\documentclass[english,11pt,oneside]{article}
\pdfoutput=1
\usepackage{amsmath}
\usepackage{amsfonts}
\usepackage{amssymb}
\usepackage{graphicx, rotating}
\usepackage{epstopdf}
\usepackage{epsfig}
\usepackage{latexsym}
\usepackage{multirow}
\usepackage{color}
\usepackage{slashed}
\usepackage{pifont}
\usepackage[top=2.8 cm, bottom=2.8 cm, left=3 cm, right=3 cm]{geometry}

\newcommand{\ba}{\begin{eqnarray}}
\newcommand{\ea}{\end{eqnarray}}
\newcommand{\no}{\nonumber}

\usepackage{hyperref}
\hypersetup{
 colorlinks = true,
 linkcolor  = red,
 citecolor={blue}
}

\usepackage{bbold}
\newcommand{\newc}{\newcommand}

\newc{\lcal}{\int {\cal L}dt}
\newc{\ie}{{\it i.e.}}          
\newc{\etal}{{\it et al.}}
\newc{\eg}{{\it e.g.}}          
\newc{\kev}{\hbox{\rm\,keV}}            
\newc{\mev}{\hbox{\rm\,MeV}}            
\newc{\gev}{\hbox{\rm\,GeV}}            
\newc{\tev}{\hbox{\rm\,TeV}}
\newc{\xpb}{\hbox{\rm\, pb}}
\newc{\xfb}{\hbox{\rm\, fb}}

\def\bea{\begin{eqnarray}}
\def\eea{\end{eqnarray}}
\def\beq{\begin{equation}}
\def\eeq{\end{equation}}

\def\lsim{\mathrel{\rlap{\lower3pt\hbox{\hskip0pt$\sim$}}
   \raise1pt\hbox{$<$}}}         %less than or approx. symbol
\def\gsim{\mathrel{\rlap{\lower4pt\hbox{\hskip1pt$\sim$}}
   \raise1pt\hbox{$>$}}}         %greater than or approx. symbol

\begin{document}

\title{Electroweak Baryogenesis above the Electroweak Scale}
\author{Alfredo Glioti~\footnote{alfredo.glioti@epfl.ch}~~~~Riccardo Rattazzi~\footnote{riccardo.rattazzi@epfl.ch}~~~~Luca Vecchi~\footnote{luca.vecchi@epfl.ch}\\
{\small\emph{Theoretical Particle Physics Laboratory, Institute of Physics, EPFL, Lausanne, Switzerland}}}
\date{}
\maketitle

\begin{abstract}

Conventional scenarios of electroweak (EW) baryogenesis are strongly constrained by  experimental searches for CP violation beyond the SM. We propose an alternative scenario where
 the EW phase transition and baryogenesis 
occur at temperatures of the order of a new physics threshold $\Lambda$ far above the Fermi scale, say, in the $100-1000$ TeV range. This way the needed new sources of CP-violation, together with possible associated flavor-violating effects, decouple from low energy observables. The key ingredient is a new CP- and flavor-conserving sector at the Fermi scale that ensures the EW symmetry remains broken and sphalerons suppressed at all temperatures below $\Lambda$. 

We analyze a minimal incarnation based on a linear $O(N)$ model. We identify a specific large-$N$ limit where  the effects of the  new sector are vanishingly small at zero temperature  while being significant at 
 finite temperature. This crucially helps the construction of realistic models. 
A number of accidental factors, ultimately related to the size of the relevant  SM couplings, force $N$  to be above $\sim 100$. Such a large $N$ may seem bizarre, but it does not affect the simplicity of the model and in fact it allows us to carry out a consistent re-summation of the leading contributions to the thermal effective potential.
Extensions of the SM Higgs sector can  be compatible with smaller values $N\sim 20-30$. 

{Collider signatures are all parametrically suppressed by inverse powers of $N$ and may be challenging to probe, but  present constraints from direct dark matter searches cannot be accommodated in the minimal model. We  discuss  various extensions that satisfy all current bounds. One of these  involves a new gauge force confining at scales between $\sim1$ GeV and the weak scale.}

\end{abstract}

\newpage

{
	\hypersetup{linkcolor=black}
	\tableofcontents
}

\section{Introduction}

The existence of a non-vanishing baryon number density in the universe is a fact of life and a mystery of  physics.
According to our present understanding of the early universe, such density cannot be simply accounted for by initial conditions, as any original density would have been completely diluted by inflation. Dynamics during the Big Bang era must be responsible for  the observed  ratio of baryon number  over entropy $\eta _b\equiv n_b/s\simeq 10^{-{10}}$. The conditions such dynamics should satisfy were famously spelled out by Sakharov a long time ago. The first condition, concerning the existence
of baryon number violating interactions, is satisfied by the Standard Model (SM). This happens in a remarkable way: on the one hand, at low temperature, baryon number emerges as an accidental symmetry whose violation is
negligibly small; on the other hand, at temperatures above the weak scale, fast baryon number violation occurs through sphaleron processes. The other two conditions, concerning CP-violation and departure from thermodynamic equilibrium, are however not satisfied by the SM. The  reasons for that are more quantitative than structural. Indeed the SM is endowed with CP violation 
in the CKM matrix, but that turns out to be insufficient, in view of the small value of quark masses and mixings. The SM, given electroweak symmetry breaking, could also, in principle, experience an epoch of departure from equilibrium  when transiting from a high temperature symmetric phase to a low temperature non-symmetric phase. However, given its parameters, in particular given the Higgs mass, such transition is known to be a smooth second order crossover. Baryogenesis thus requires new physics.

A potentially testable option is having baryogenesis occurring at the electroweak phase transition~\cite{Dimopoulos:1978kv}\cite{Kuzmin:1985mm}\cite{Cohen:1990py}. A realistic model of electroweak (EW) scale baryogenesis should involve new states so as to  provide new sources of 
CP violation and a first order electroweak  phase transition. The new states should range from $\sim 100$ GeV to several
hundreds GeV, with the upper range attainable only in the presence of somewhat strong couplings (see for instance \cite{Grojean:2004xa}). While this is interesting in view of a full test of this scenario, it also implies, given the lack of clear evidence so far of new physics in direct and indirect searches, that the models are already under pressure. In particular one major source of pressure comes from the need for new CP-violating phases: on the one hand these are directly constrained by the searches for EDMs
of elementary particles, %in particular the electron, 
on the other their presence is typically associated to new
sources of flavor violation, with the corresponding strong bounds. To be more quantitative, a new physics sector characterized by a mass scale $\Lambda$ and CP-odd phases $\phi_{\rm CP}$ is expected to generate an electron EDM either at 1-loop (when direct couplings of order $g$ to the SM fermions are present), or 2-loop (when mainly couplings $g$ to the bosonic sector of the SM exist). Taking $g$ of order the EW gauge coupling for illustration, in these two cases we estimate:
\ba
\frac{|d_e|}{e}=
\begin{cases}
\sin\phi_{\rm CP}\frac{g^2}{16\pi^2}\frac{m_e}{\Lambda^2}\sim10^{-29}{\rm cm}\times\sin\phi_{\rm CP}\left(\frac{50~{\rm TeV}}{\Lambda}\right)^2 & {\rm1-loop}\\
\sin\phi_{\rm CP}\left(\frac{g^2}{16\pi^2}\right)^2\frac{m_e}{\Lambda^2}\sim10^{-29}{\rm cm}\times\sin\phi_{\rm CP}\left(\frac{2.5~{\rm TeV}}{\Lambda}\right)^2 & {\rm2-loop}~({\rm Zee-Barr})
\end{cases}
\ea
The recent improved bound on the electron EDM $|d_e|<1.1\times10^{-29}e~{\rm cm}$~\cite{ACME} then suggests that $\Lambda$ should be well above the weak scale if $\sin\phi_{\rm CP}\sim1$. Scenarios with suppressed CP-phases may allow the new physics to stay in the hundreds of GeV, but in those cases achieving a sizable baryon asymmetry is rather difficult (see, e.g., \cite{Cirigliano:2009yd}). Even ignoring possible associated flavor-changing effects we must conclude that realizing conventional scenarios of EW baryogenesis is at best extremely challenging. Clever flavor model-building provides one way to limit this pressure, nonchalance provides another.
The goal of this paper is to study an alternative scenario, where all constraints from flavor and CP violation are structurally eliminated.

Our scenario is described as follows. Up to some high scale $\Lambda \gg m_W$, say $\Lambda = 100-1000$ TeV, the only sources of flavor and CP violation are the SM Yukawa couplings. One should think of $\Lambda$ as the scale of flavor, at which new sizable sources of CP and flavor violation become active, without significantly affecting low energy observables.  In view of that, $\Lambda$ seems the natural scale where to realize baryogenesis. Our idea is then simply that the electroweak phase transition happens at $T\sim \Lambda$ rather than at $T\sim m_W$ and that moreover such a transition is first order with consequent departure from thermodynamic equilibrium and generation of a baryon asymmetry. In order for such asymmetry to be kept unsuppressed until the present day, the EW symmetry
must remain broken at all temperatures below $\sim \Lambda$, so as to avoid wash-out from sphaleron processes. In the SM, thermal effects, dominated by the top quark, are known to restore the EW symmetry at a temperature $T_c\simeq 160$ GeV (see e.g.~\cite{DOnofrio:2014rug} for a recent calculation). In order to realize this scenario there should therefore exist new degrees of freedom in the $100$ GeV range
and coupled to the Higgs so as to prevent electroweak symmetry restoration at temperatures above the weak scale and below $\Lambda$. The simplest option for such states is given by a set of $N$ SM neutral scalar fields $S_i$ bilinearly coupled to the Higgs.
 As we shall discuss in detail, there exists a minimal value of $N$ where the Higgs vacuum expectation value (VEV) at finite temperature is big enough to suppress sphaleron processes at $T< \Lambda$ while preserving perturbativity.
For reasons that emerge combining the structural consistency of the model with the significant impact of the top Yukawa on the Higgs potential, this minimal $N$ turns out to be quite large, $N\gtrsim100$. Such large numbers, however, do not imply complexity in the structure of the model: the scalars $S_i$ could fit in a single multiplet of a global or local symmetry and the Lagrangian could be described by a handful of couplings. Indeed we shall also make a simple remark showing there exists a scaling region for the couplings and for $N$, where 
finite temperature effects are large while zero temperature ones, like collider production rates or corrections to electroweak precision quantities, decouple with inverse powers of $N$.

The main goal of this paper is to study in detail the dynamics  at temperatures below $\Lambda$. The $S_i$ will have a mass in the $100$ GeV range, so we must ensure they serve our purpose (symmetry non-restoration)  compatibly with all theoretical
and phenomenological constraints. In particular one issue concerns their problematic relic density, which we shall eliminate by introducing additional interactions which can in principle be rather weak. 

For what concerns the EW phase transition dynamics at the scale $\Lambda$, there is great freedom as there are no significant constraints on model-building at scales $\Lambda\sim100-1000$ TeV from low energy phenomenology. That was indeed the goal of the whole construction. In view of that we shall limit ourselves to a simple sketch of possible scenarios at the scale $\Lambda$. 

Our paper is organized as follows. In Section \ref{sec:sphaleron} we identify a sufficient (and largely model-independent) condition to ensure the primordial $B+L$ asymmetry is not washed-out by sphaleron processes at $T<\Lambda$. Our $O(N)$ model is introduced in Section \ref{sec:model}, where its domain of validity as a perturbative effective field theory (EFT) is established. Especially important for the reminder of the paper is how the perturbativity and stability conditions scale with $N\gg1$. We will see that large-$N$ sectors have the ability to qualitatively impact the finite-temperature dynamics while remaining essentially invisible around the vacuum. In Section \ref{firstapprox} we take a first look at the finite $T$ behavior of our model and  roughly assess how large $N$ needs to be in order to ensure EW symmetry non-restoration. The need for a refined analysis is emphasized in Section \ref{sec:T}. Here we carefully assess the perturbative regime of the finite $T$ version of our scenario and discuss some computational  subtleties that characterize our class of large-$N$ models. A detailed discussion of the effective potential at finite $T$ is presented in Section \ref{sec:effV}. A large-$N$ technique is implemented to reliably take into account the most important contributions. A numerical analysis is then used to identify the allowed parameter space of the model. The main phenomenological constraints on the minimal $O(N)$ scenario are presented in Section \ref{sec:pheno}. All direct and indirect signatures of $S_i$ at colliders are suppressed by powers of $N$ and may be hard to see. Simultaneously, the minimal model predicts $N$-enhanced signals at direct dark matter experiments and is therefore incompatible with current data. We thus discuss various possible extensions that eliminate the problem. A subset of these extensions also features a realistic dark matter candidate.
In particular in subsection \ref{sec:darkGlue} we focus on a scenario where the $O(N)$ symmetry is partially gauged and where dark matter can plausibly be made of bound states. This is a novel dark matter scenario  that crucially combines compositeness and large-$N$.
In view of its  intricate phenomenology, subsection  \ref{sec:darkGlue}  somewhat blew out of proportions: in spite of its interest this section may be skipped in a first reading. In Section~\ref{sec:alter} we sketch possible variants of our scenario. These include models with different symmetry structure but a comparable or even larger number of degrees of freedom than in the model discussed in detail in the paper. However we also point out that a large $N$ is not strictly necessary to our program if we introduce additional EW-charged scalars. In Section \ref{sec:100TeV} we show that there exists no structural obstruction to the construction of complete models for EW baryogenesis. We illustrate this by briefly discussing two possible scenarios of new physics at the flavor scale $\Lambda$, one strongly- and one weakly-coupled. We present our conclusions in Section~\ref{sec:discussion}.

\section{The low energy sector}

In this section, assuming baryogenesis takes place via a first order electroweak phase transition at $T\sim \Lambda \gg m_W$ (see Section~\ref{sec:100TeV} for some models for the phase transition), we shall study the conditions on the effective theory at $E\ll \Lambda$ for the baryon asymmetry to survive. The crucial request 
concerns the suppression of sphaleron wash-out. After quantifying this request in the next subsection, we will introduce a specific model and study it in detail.

\subsection{Avoiding sphaleron washout}
\label{sec:sphaleron}

Our basic assumption is that the baryon asymmetry is generated by the anomalous electroweak baryon number violation during the electroweak phase transition at $T\sim \Lambda$. More precisely, this implies the asymmetry arises 
along the direction $B+L$, which is anomalously broken by the electroweak interactions, while it vanishes in the exactly conserved direction $B-L$. Because of that we must ensure that the anomalous $B+L$-violating processes, the so-called sphalerons, remain inactive at $T<\Lambda$, as they would otherwise wash-out this very same combination.~\footnote{This issue of course does not arise in other scenarios of high scale baryogenesis, such as leptogenesis, where the asymmetry is generated along the non-anomalous $B-L$ direction.}

The rate of variation of the baryon number density normalized to the entropy density $\Gamma_b(T) \equiv -\frac{d}{dt}\ln \eta_b$
crucially depends on the weak coupling $\alpha_w=g^2/4\pi$, on the temperature $T$, and on the temperature-dependent Higgs vacuum expectation value (VEV) $h(T)$.~\footnote{Throughout the paper we adopt the convention $h(0)\equiv v=246$ GeV, for which $m_W^2=g^2h^2/4$ and $m_t^2=y_t^2h^2/2$.} 
One can distinguish three qualitatively different regimes, which respectively correspond to high, intermediate and low temperature:
\ba
{\rm high~T}: && T\gtrsim \frac{gh(T)}{\alpha_w}= \frac{4\pi}{g} h(T)\\\no
{\rm inter.~T}: && \frac{g}{4\pi}h(T)\lesssim T\lesssim \frac{4\pi}{g}h(T)\\\no
{\rm low~T}:&&T\lesssim \frac{g}{4\pi}h(T).
\ea
These ranges are essentially determined by the $T$-dependent vector boson mass $m_W\sim g h(T)$, and by the energy of the sphaleron configuration 
 \beq
 E_{\rm sph}(T)\equiv B \frac{gh(T)}{\alpha_w},
 \eeq
where $B=B(\lambda_H/g^2,g'^2/g^2)$ is a number of order unity. According to the zero-temperature analysis of refs.~\cite{Klinkhamer:1984di}\cite{Kleihaus:1991ks}, $B\sim1.9$. 

In the high temperature regime the EW-symmetry is effectively unbroken: the sphaleron barrier is overcome by thermal fluctuations, the rate is $\Gamma_{\rm b}\sim20\alpha_w^5T\gg H$~\cite{Moore:1997sn}\cite{DOnofrio:2014rug} and any asymmetry would be washed-out in a fraction of a Hubble time. On the other hand, the low temperature regime approaches the zero temperature result, where $B+L$-violating effects are $\propto e^{-2\pi/\alpha_w}$ and can be safely neglected under all relevant circumstances. Finally, in the intermediate regime one finds
\ba\label{sphrate}
\Gamma_b(T)\sim 0.1\, T\left(\frac{gh}{T}\right)^7\left(\frac{4\pi}{\alpha_w}\right)^3\,e^{-{E_{\rm sph}}/{T}}\,,
\ea
where  to estimate the prefactor we used the results of~\cite{Arnold:1987mh} with $\kappa\sim1$ and $\omega_-\sim gh$. The reliability of this estimate of the actual sphaleron rate can be assessed observing that \eqref{sphrate} fits the numerical results of~\cite{DOnofrio:2014rug} quite well provided $B\to B_{\rm fit}\simeq2.2-2.3$.~\footnote{We thank M. Shaposhnikov for suggesting this fit.} We thus see that finite $T$ effects modify the rate \eqref{sphrate} appreciably. In what follows we will still use \eqref{sphrate}, but will adopt the conservative value $B=1.8-2.1$.

%In this sense our choice $B=1.8-2.1$ is perhaps a bit conservative.

As we will see shortly, in order to preserve the baryon asymmetry generated at the phase transition, it is sufficient to ensure that, when  the universe was  in the range of temperatures $m_W\lesssim T \lesssim \Lambda$, the rate in eq.(\ref{sphrate}) was small enough. One can easily see that the constraint on $h(T)/T$ is strongest at the latest times, that is the lowest $T$'s in the range of interest. Indeed the preservation of the baryon asymmetry essentially gives the constraint $\int \Gamma_b dt \lesssim 1$, which by using Hubble law  can be conveniently written as the  $dT$ integral
\beq\label{gammab}
\int \Gamma_b(T)\frac{M_P}{\sqrt {g_*}}\frac{dT}{T^3}=\int F(gh(T)/T)\frac{M_P}{\sqrt{g_*} T}\frac{dT}{T}\lesssim 1
\eeq
where $F(x)\propto x^7e^{-Bx/\alpha_w}$ --- implicitly defined by eqs.(\ref{sphrate}) and \eqref{gammab} --- goes monotonically to $0$ as $x\to 0$.
The strongest constraint on $h(T)/T$ is dominated by the region where $T$ is of order the weak scale and reads
\ba\label{strongPT}
\frac{h(T)}{T}\gtrsim1.2-1.5,
\ea
with the upper (lower) value corresponding to $B\simeq1.8\,(2.1)$. %The latter condition is numerically equivalent to $h(T)/T\gtrsim1.2-1.5$. 
As promised at the beginning of the paragraph, it is now apparent that requiring \eqref{strongPT} for all $T< \Lambda$ automatically prevents us from entering the high $T$ regime, where washout is unsuppressed. We therefore conclude that eq.~(\ref{strongPT}) represents a sufficient condition to avoid washout. Note also that in the models we shall study $h(T)/{T}$ is minimized at around the weak scale, and therefore the argument leading to the numerical values shown in eq.~(\ref{strongPT}) is justified.

\subsection{A model}
\label{sec:model}

In order to guarantee the condition on $h(T)/{T}$ discussed in the previous section we must suitably modify the Higgs dynamics in the energy range between $m_W$ and $\Lambda$. A simple model realizing that is constructed by adding to the SM a multiplet
of scalars $S_i$ transforming as the fundamental representation of a novel $SO(N)$ symmetry. Other options, based on different symmetries (possibly discrete) and multiplet content will be mentioned in Section~\ref{sec:alter}. Besides the gauge and Yukawa couplings of the SM, the resulting model is defined by the scalar potential
\ba\label{mod}
V={m_H^2}H^\dagger H+{\lambda_H}(H^\dagger H)^2+\frac{m_S^2}{2}S^2+\frac{\lambda_S}{4}(S^2)^2+{\lambda_{HS}}S^2H^\dagger H,
\ea
which is the most general one based on symmetries and renormalizability. 
Here the SM and $O(N)$ indices have been suppressed for brevity. We will assume 
\ba\label{muS}
m_H^2<0~~~~~~~~m_S^2>\frac{\lambda_{HS}}{\lambda_H}m_H^2
\ea
to ensure that in vacuum the Higgs has a non-vanishing expectation value whereas $\langle S\rangle=0$.

The basic idea, which is not new~\footnote{Symmetry non-restoration was first realized within a 2-scalar model (essentially the small $N$ version of the one we adopt below) by S. Weinberg~\cite{Weinberg:1974hy}. The same picture was subsequently applied in different contexts by other authors. Ref. \cite{Mohapatra:1979vr} implemented it in models with spontaneous CP violation; ref. \cite{Salomonson:1984rh} proposed that non-restoration might be used to avoid the monopole problem of GUTs, and the same application has been suggested in \cite{Dvali:1995cj}. Spontaneous $B$ violation at finite $T$ was used in a model for baryogenesis via decays in \cite{Dodelson:1989ii}, and a period of temporary color $SU(3)$ breaking has been employed in \cite{Ramsey-Musolf:2017tgh} in a model for EW baryogenesis. More recently, the idea of EW symmetry non-restoration has been put forward in \cite{Meade:2018saz} and applied in the same context as ours in \cite{Baldes:2018nel}.}

but we believe we shall explore from a novel perspective, is this: a negative off-diagonal quartic, $\lambda_{HS}<0$, provides $H$ with a negative Debye mass at finite temperature, so that $\langle H\rangle \not =0$ even at $T\gg m_W$. Of course this makes sense as long as  
the potential is bounded from below and as long as all the couplings remain  perturbative in a range of energies above the weak scale.

At tree level, the conditions for a potential bounded from below are easily found to be
\ba\label{stab}
\lambda_H,~\lambda_S>0~~~~~~~\lambda_{HS}>-\sqrt{\lambda_H\lambda_S}
\ea
showing there exists a window $-\sqrt{\lambda_H\lambda_S}<\lambda_{HS}<0$ where a negative $\lambda_{HS}$ can in principle
achieve our goal of symmetry non-restoration. In order to ensure the absence of instabilities within the EFT description below the scale $\Lambda$, the above conditions will have to be satisfied by  the running couplings at all RG scales $\mu \lesssim \Lambda$.

The  conditions ensuring our construction makes sense as a weakly coupled EFT can be established  by a simple diagrammatic analysis to be
\ba\label{pert} 
\epsilon_H&\equiv&\frac{6\lambda_H }{16\pi^2}\ll1,\\\no
\epsilon_S&\equiv&\frac{\lambda_SN}{16\pi^2}\ll1,\\\no
|\epsilon_{HS}|&\equiv&\frac{2|\lambda_{HS}|\sqrt{N}}{16\pi^2}\ll1,
\ea
where we have allowed for the possibility that $N\gg1$ be counted in. One direct way to derive these constraints is to consider  $s$-wave $2\to 2$ scattering in singlet states
\beq
|HH\rangle \equiv \frac{1}{\sqrt 4}\left (|H_1H_1\rangle+\dots+|H_4H_4\rangle\right )\qquad\qquad |SS\rangle \equiv \frac{1}{\sqrt N}\left (|S_1S_1\rangle+\dots+|S_NS_N\rangle\right)\, .
\eeq
In the 2-dimensional Hilbert subspace with basis given by the $\ell=0$ singlet states $\{|HH\rangle,\,|SS\rangle\} $ the $S$-matrix is 
\beq
S\equiv e^{2i\delta}=\mathbb{1}+\frac{i}{8\pi} {\cal T}+\dots
\eeq
with
\beq
{\cal T}=\left ( \begin{array}{cc} 6\lambda_H& 2\lambda_{HS}{\sqrt N}\\ 2\lambda_{HS}{\sqrt N}& \lambda_S(N+2)\end{array}\right )
\label{Tmatrix}
\eeq
and with the dots corresponding to terms beyond the tree approximation. 
By the above equation we deduce
\beq
\delta = \frac{1}{16\pi}{\cal T}+\dots
\eeq
and requiring,  in the spirit of naive dimensional analysis (NDA), that all the eigenvalues of $\delta$ be $\lesssim \pi$, we derive constraints that are parametrically equivalent to eq.~(\ref{pert}). The rule according to which $\sqrt N$ appears in eq.~(\ref{Tmatrix}) is easily established: any $SS$ singlet, either initial or final, counts $O(\sqrt N)$, while $HH$ counts $O(1)$.
One is easily convinced that  scattering in non-singlet channels gives weaker bounds, as the amplitudes are not enhanced by powers of $N$. Alternatively, the same constraints (\ref{pert}) can be derived by considering the $\beta$-functions for $(\lambda_H,\lambda_S,\lambda_{HS})$, see \eqref{betafunctions} for the 1-loop approximation, and requiring that the relative change of any of them is less than $O(1)$ over one $e$-folding of RG evolution. Notice that the window of negative $\lambda_{HS}$ allowed by stability, eq.~(\ref{stab}), is described by  $\epsilon_{HS}^2\leq (2/3)\epsilon_H\epsilon_S$. Therefore, perturbativity of the diagonal quartics  $\lambda_{H,S}$ plus stability guarantees perturbativity of the off-diagonal quartic $\lambda_{HS}$.

\begin{figure}[t]
\begin{center}
\includegraphics[scale=1]{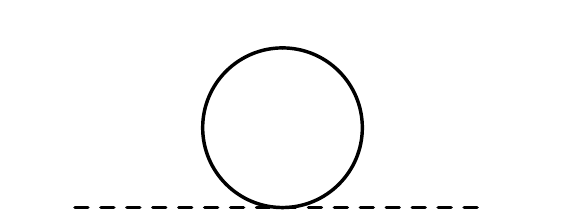}~~~~~~~~
\includegraphics[scale=1]{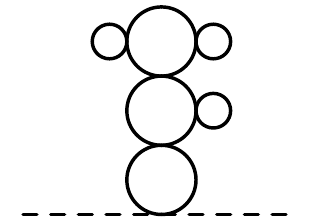}
\caption{LEFT: leading perturbative contribution to the Higgs thermal mass. The solid (dashed) lines correspond to $S$ $(H)$. RIGHT:  a sample of diagrams re-summed by the large-$N$ technique adopted in Section~\ref{sec:effV}.
}\label{fig:tadpole}
\end{center}
\end{figure}

With eq.~(\ref{pert}) satisfied, our model is a well defined weakly coupled EFT. Yet, remarkably, in the large-$N$ limit there remains one class of calculable quantum effects that is not suppressed by the $\epsilon$'s. These have to do with the renormalization of the mass of $H$  induced by the $S$-loop tadpole diagram in the left panel of Fig. \ref{fig:tadpole}. For instance, in vacuum this diagram gives a contribution to the RG evolution of $m_H^2$:
\beq
\mu\frac{d}{d\mu} m_H^2 =\frac{\lambda_{HS}N}{8\pi^2}m_S^2\equiv \epsilon_{HS}\sqrt N m_S^2,
\label{RGmass}
\eeq
implying that for $\sqrt N\gtrsim 1/|\epsilon_{HS}|$ there remains a finite effect,  for arbitrarily small $|\epsilon_{HS}|$. 
When  compactifying some spacetime directions the same diagram contributes finite mass corrections of Casimir type. In particular, by considering the system at finite temperature and working with compactified euclidean time,  this diagram provides a negative contribution to the thermal mass~\footnote{Here we consider $m_S\ll T$ and neglected refinements  needed when $S$ itself acquires a sizable thermal mass and which we shall discuss in detail later on.}
\beq\label{thermalhs}
\delta m_H^2=\frac{\lambda_{HS} N}{12} T^2\,=\frac{2\pi^2}{3} \epsilon_{HS}\sqrt N T^2<0.
\eeq
This result implies a very interesting property of this simple system at large-$N$: in the scaling limit $\epsilon_{H,S,HS}\to0$, $N\to \infty$ with $\epsilon_{HS}\sqrt N$ fixed, the theory is free, in particular the $S$-matrix is trivial, yet at finite temperature a finite contribution to the $H$ thermal mass survives. For instance, in the case  $m_H^2>0$, where the symmetry is unbroken in the vacuum, an infinitesimally small $\epsilon_{HS}<0$ at sufficiently large $N$ can still generate a finite negative thermal mass and thus trigger symmetry breaking at finite temperature. At large-$N$ we can thus have a free theory, which when heated up undergoes a phase transition. While this may seem paradoxical, one should consider that in order to heat up a system of $N\to \infty$ fields one must provide an energy density that diverges in the same way: the phase transition cannot occur at finite energy density.

%%%%%%%%%%%%%%%%%%
%%%%%%%%%%%%%%%%%%
%%%%%%%%%%%%%%%%%%
%%%%%%%%%%%%%%%%%%

The crucial implication of eq.~(\ref{thermalhs}), is that for sufficiently large $N$, eq.~(\ref{strongPT}) can be ensured at all $T<\Lambda$ with $\epsilon_{HS}$ and $\epsilon_S$ small enough not to significantly perturb the SM dynamics and compatibly with stability, $\epsilon_{HS}^2<(2/3)\epsilon_H\epsilon_S$. The main question concerns the minimal value of $N$ for which all constraints are met. As we shall discuss in detail in the next subsections, for a combinations of factors, the needed value of $N$ is quite large.

Before proceeding we would however like to make a few  comments concerning naturalness. 
In our construction 
there is clearly no protection mechanism for the hierarchical separation between $m_H, m_S$ and $\Lambda$. We do not want to try and further complicate the model to make this separation natural. Indeed, in the case of conventional approaches to the hierarchy, i.e. supersymmetry or compositeness, that would almost unavoidably force us to have to deal with the problem of new sources of flavor violation at the weak scale, which is what we wanted to avoid in the first place. One perhaps tenable
 perspective is that  this separation of scales has an anthropic origin. The Higgs mass $m_H$ could be pegged to its value by the atomic principle \cite{Agrawal:1997gf}, that is by the necessity for $m_W$ to be not too much above $\Lambda_{\rm QCD}$ in order to ensure nuclear and chemical complexity. The $S$ mass $m_S$, on the other hand, could be pegged to the weak scale by the request of a sufficiently baryon-rich universe: $m_S>  m_W$  would cause electroweak symmetry restoration at $m_W\lesssim T\lesssim m_S$ with corresponding sphaleron washout of the baryon asymmetry created at the scale $\Lambda$. Of course, we are aware of the weakness of such anthropic reasoning: once one invokes a multiverse of options it is well possible the baryon asymmetry can be generated by some other mechanism, for instance by leptogenesis. Still our simple argument does not seem fully implausible to us.~\footnote{In \cite{ArkaniHamed:2005yv}\cite{Senatore:2005ch} anthropic arguments were similarly invoked to motivate the existence of a weak scale sector with all the ingredients necessary to realize EW baryogenesis.}
 
On the quantitative side, notice that,  given the physical cut-off $\Lambda$, we expect a finite correction to $m_H^2$ 
\beq
\delta m_H^2\sim \frac{\lambda_{HS} N}{8\pi^2} \Lambda^2\,=\epsilon_{HS}\sqrt N \Lambda^2.
 \eeq
 Therefore the $\sqrt N$ enhancement of the $S$-tadpole, which is essential for the vacuum dynamics at finite $T$,   implies a corresponding increase in the needed tuning.  In practice we shall need $|\lambda_{HS}| N\sim 8$, so that the needed tuning is only a factor of a few worse than the one due to the top quark contribution.

\subsection{Thermal vacuum dynamics in first approximation}
\label{firstapprox}
Here we would like to give a first rough idea of the range of parameters where the model is compatible with our scenario for baryogenesis. In order to do so we approximate the thermal effective potential by the tree level potential with quartic couplings renormalized at a scale $\sim T$ and supplemented by the leading 1-loop thermal corrections to the scalar masses. In the next two sections we will perform a more accurate computation. 

Now, according to the above approximation, the thermal potential is given by eq.~(\ref{mod}) with the scalar masses replaced by
\ba\label{zeromodemass1}
m_H^2&\to &m_H^2(T)=m_H^2+ \left[\frac{N}{12}\lambda_{HS}+\frac{1}{2}\lambda_H +\frac{3}{16}g^2+\frac{1}{16}g'^2+\frac{1}{4}y_t^2\right]{T^2}\\\no
m_S^2&\to&m_S^2(T)=m_S^2+ \left[\frac{(2+N)}{12}\lambda_S+\frac{1}{3}\lambda_{HS}\right]{T^2},
\ea
where all couplings are understood to be evaluated at an RG scale $\sim T$. In the above equation we have included the thermal corrections induced by $S$ as well as those by the leading SM couplings. Notice that the negative $\lambda_{HS}$ contribution to $m_S^2(T)$ is not  $N$-enhanced, so that in the scaling limit $N\to \infty$ with $\lambda_SN$ and 
$\lambda_{HS}N$ fixed it can be neglected. Moreover, it is easy to see that \eqref{muS} is satisfied by the thermal masses in eq.~\eqref{zeromodemass1}, implying that $S$ does not acquire a VEV. On the other hand, we have $\langle H\rangle=(0, h/\sqrt{2})^t$, with 
\ba
\label{honTrough}
\frac{h^2(T)}{T^2}&=&-\frac{m_H^2(T)}{ T^2 \lambda_H}\simeq-\left[\frac{N}{12}\lambda_{HS}+\frac{1}{2}\lambda_H +\frac{3}{16}g^2+\frac{1}{16}g'^2+\frac{1}{4}y_t^2\right]\frac{1}{\lambda_H}\\\no
&\equiv&\frac{\frac{N}{12}|\lambda_{HS}(T)|-A(T)}{\lambda_H(T)}
\ea
 where we have grouped all the SM contributions into the quantity $A(T)>0$. We conservatively neglected the zero temperature mass; being negative, it helps making $h/T$ bigger but also becomes quickly negligible as soon at $T\gtrsim 100$ GeV.

Now, given eq.~(\ref{honTrough}), the requirement $h(T)/T\gtrsim 1.2$ in eq.~(\ref{strongPT}) translates into a lower bound $|\lambda_{HS}(T)|N\gtrsim 12[A(T)+(1.2)^2\lambda_H(T)]$, which we shall impose for all $T< \Lambda$. This bound is strongest at $T\sim 100-1000$ GeV where it reads $|\lambda_{HS}| N\gtrsim8$, since the SM couplings are larger there. Moreover, we must simultaneously impose the stability requirement $\lambda_{HS}^2(\mu)\leq \lambda_H(\mu)\lambda_S(\mu)$ at any scale $\mu <\Lambda$, to ensure our effective field theory makes sense. Combining these two constraints we obtain
\bea\label{Nrough}
N&\geq&\frac{[N\lambda_{HS}(\mu)]^2}{\lambda_H(\mu)[N\lambda_S(\mu)]}=\frac{[N\lambda_{HS}(T)]^2}{\lambda_H(\mu)[N\lambda_S(\mu)]}\left (\frac{\lambda_{HS}(\mu)}{\lambda_{HS}(T)}\right )^2\\\no
&\gtrsim&\frac{[12]^2\left [A(T)+(1.2)^2 \lambda_H(T)\right ]^2}{\lambda_H(\mu)[16\pi^2\epsilon_S(\mu)]}\left (\frac{\lambda_{HS}(\mu)}{\lambda_{HS}(T)}\right )^2\\\no
&\sim&
800 \left (\frac{0.01}{\epsilon_S(\mu)}\right )\frac{0.04}{\lambda_H(\mu)}\left (\frac{\lambda_{HS}(\mu)}{\lambda_{HS}(\rm TeV)}\right )^2\gtrsim 800 \left (\frac{0.01}{\epsilon_S(\Lambda)}\right).
\eea
The right hand side in the second line of Eq. \eqref{Nrough} decreases with $T$ as the SM couplings  and $|\lambda_{HS}|$ respectively decrease and increase with $T$. Moreover that same expression increases with $\mu$, with the quantitatively most important effect being associated with the decrease in $\lambda_H$ by a factor of about $3-4$ when running between the weak scale and $100-1000$ TeV. The final bound on $N$ is therefore maximized at $T\sim$ TeV and for the largest $\mu\sim \Lambda$, which in our picture could be $100$ to $1000$ TeV. We conclude that unless $\epsilon_S(\Lambda)$ is rather strong, $N$ must  be in the hundreds. Notice also that the limiting values $N\sim 800$ and $|\lambda_{HS}| N\sim 8$ consistently correspond to a rather weak coupling $|\epsilon_{HS}|\sim 0.004$. 

There are two ways in which the estimate \eqref{Nrough} is inaccurate. The first is that in the region where $h/T\sim 1$ the thermal masses of the Higgs, $W^\pm$, $Z$, and top-quark are not negligible compared to the temperature, in which case the thermal loop correction to the Higgs potential starts being affected by Boltzmann suppressions causing a deviation from the simple expressions used in eq.~(\ref{zeromodemass1}). This effect tends to reduce $A(T)$ by a few $10\%$ percent thus allowing slightly smaller values of $N$. The second reason why \eqref{Nrough} is not fully accurate is that higher order corrections become significant already for $\epsilon_S\sim 0.01$. That is related to the well-known poor convergence of thermal perturbation theory. Whether or not these latter effects can help decrease the minimum value of $N$ is an intricate question that necessitates the careful analysis presented in the following sections. In the next subsection we begin with a qualitative assessment of these issues.

\subsection{3D or not 3D?}
\label{sec:T}
In order to set the stage for the refinement of the estimate of the previous section,
we will here first discuss the issue of perturbation theory, which is the most important and subtle, and later address how a sizable  $h/T$ affects \eqref{Nrough}. The former issue was originally discussed many years ago in the context of gauge theories, see e.g.~\cite{Ginsparg:1980ef}\cite{Linde:1980ts}\cite{Gross:1980br}\cite{Appelquist:1981vg}, while the second is a novel feature of the present framework.

The poorer convergence properties  of perturbation theory in thermal field theory compared to QFT at zero temperature, stems from two joint 
facts. The first is the presence  of light bosonic degrees of freedom (associated to the Matsubara zero modes) in the 3D effective  theory below the scale $\sim\pi T$. 
 The second is the IR relevance  of their interactions, i.e. the presence of couplings of positive mass dimension that  become strong at sufficiently low momenta.
These facts can indeed lead to a complete breakdown of perturbation theory at finite temperature even for perfectly weakly coupled QFTs. One example of that is given by massless non-abelian gauge theories, which become strongly coupled at a scale roughly of the order of their effective 3D coupling $\propto g_3^2=g_4^2T$. This for instance implies the well know result that the free energy of hot QCD does not admit a series expansion in $g_4^2$ beyond the third order. Another example is given by systems in the neighborhood of a phase transition where some scalar is tuned to be light. The resulting 3D strong coupling can  make it difficult to assess  the order of the phase transition. Nevertheless, even without going to such limiting  situations, it is always the case that perturbation theory converges more slowly
at finite temperature.

In order to better appreciate this issue let us consider the simple case of $\lambda \varphi^4$. The dimensionless 4D quartic  coupling $\lambda$ matches to a dimension 1 coupling $\lambda_{eff}= \lambda T$  in the 3D effective theory at scales $p\ll\pi T$. 
In this situation, the loop expansion parameter at external momentum $p$ in the massless limit  is proportional to $\lambda_{eff}/p$
and grows strong at sufficiently small $p$. The presence of a finite mass saturates this IR growth. One can easily estimate
the strength of the loop expansion parameter by comparing, for instance, the tree and 1-loop contributions to the 4-point function at small external momentum.~\footnote{We cannot proceed as  in the previous section because in the  Euclidean 3D theory there is no $S$-matrix.  The estimate is however comparable to that implied by the $S$-matrix of the 3D Minkowskian theory that is obtained through Wick rotation.} One finds that the loop expansion is  roughly controlled by $\epsilon_{3D}=(3\lambda T)/(16\pi m_{\varphi, 0})=\epsilon_{4D}(\pi T/m_{\varphi, 0})$, as dictated by dimensional analysis. In this equation $m_{\varphi, 0}$ represents the mass of the zero mode including  of course
all thermal and quantum corrections. As we already mentioned, in the vicinity of a second order phase transition the effective mass $m_{\varphi, 0}$ can be small enough to make the IR 3D dynamics strongly coupled even for 
an arbitrarily weak 4D quartic $\lambda$. However, even away from such situation and focussing on the high temperature regime,
one finds that $\epsilon_{3D}$ is larger that its 4D counterpart. Indeed, at temperatures much larger than the 4D mass $m_\varphi$, the mass of the zero mode is dominated by thermal corrections, which at sufficiently weak coupling are well approximated by the 1-loop contribution
\beq
m_{\varphi,0}^2=\frac{\lambda}{4} T^2+ \dots
\label{thermalphi}
\eeq
so that the 3D loop expansion parameters is
\beq
\epsilon_{3D}\equiv \frac{3\lambda T}{16\pi m_{\varphi,0}}\sim \frac{3 \sqrt {\lambda}}{8\pi}=\frac{\sqrt{3}}{2}\sqrt{\epsilon_{4D}},
\label{3vs4}
\eeq
where $\epsilon_{4D}=3\lambda/(16\pi^2)$ is consistent with~\eqref{pert}. This ``square root" relation among the expansion parameter in respectively 3D and 4D is at the origin of the lower convergence rate of perturbation theory in thermal field theory. For instance, $\epsilon_{4D}=0.01$, known by practice to be well within the perturbative region, corresponds to $\epsilon_{3D}\sim 0.1$, which can easily lead to poor convergence in the presence of upward numerical ``accidents". Indeed our $\epsilon$'s
are just rough indicators of the convergence of the perturbative series: perturbation theory can be safely applied when they are significantly $\ll 1$, but not necessarily so when they are just a few times smaller than $1$. For instance the next loop order contribution within the effective 3D theory to eq. (\ref{thermalphi})~\footnote{This contribution corresponds from the point of view of 4D diagrammatics to the resummation of the so called daisy diagrams.} is of relative size $4\times \epsilon_{3D}$, which is $\sim 40\%$ for a thoroughly weakly coupled 4D theory with $\epsilon_{4D}=0.01$.
The need for applying resummation techniques in thermal field theory, is related to these simple  facts.

We can easily adapt the above discussion to our model, focusing on the scalar sector, where the parameter  $\epsilon_S$ can potentially be sizable.
 In principle one should also consider the light bosonic modes from the SM gauge sector. However, as it turns out, the overall contribution of these modes is subdominant and the associated higher order effects are thus  not very significant. Indeed the leading SM contribution is by large the one from the top quark, which we do not expect to suffer from 3D effects given the
fermionic Matsubara modes are all gapped. In view of that we shall neglect higher order effects from the gauge sector.

For what concerns the scalar sector there are two main differences with respect to the simple $\lambda\varphi^4$ case analyzed above.
The first is that ours is a multi-field model. The second is that the zero mode of the Higgs field has negative squared mass at all $T$'s, leading to symmetry breaking. Because of this second feature, the $H$ spectrum separates into the radial mode $h$ plus the (eaten) Goldstone bosons. Moreover,  in the shifted vacuum,  trilinear interactions appear. As in 3D the strength of the loop expansion parameter depends
on the IR details (the thermal masses), one should in principle perform a detailed analysis to establish the regime of perturbativity of the theory. However, and not surprisingly, if the vector boson mass (eaten Goldstones) and Higgs are roughly of the same order then the estimates are quantitatively the same as in the absence of symmetry breaking when written in terms of the physical $H$ mass. In particular, the trilinears that emerge in the broken phase do not change the overall picture qualitatively.~\footnote{The stability constraint $\lambda_{HS}^2< \lambda_H\lambda_S$ plays a role to ensure this result.} We will therefore simplify the present discussion considering the 3D effective theory for $H$ and $S$ in the absence of symmetry breaking.

The loop expansion parameters can be deduced by comparing leading and subleading contributions to  some observable. In the absence of a 3D $S$-matrix,  a surrogate of the analysis of section \ref{sec:model} is offered by the 2-point functions of linear combinations of the canonically normalized composite operators $O_S=S^2/\sqrt{N}$ and  $O_H=H^\dagger H/\sqrt{4}$. Inspecting $\langle OO\rangle$ we find that the loop expansion parameters are:
\ba\label{pertT}
%&&\epsilon^{3\rm D}_h\equiv\frac{3\lambda_H}{8\pi}\frac{T}{{m_{h,0}}}={\color{red}\epsilon_H\frac{2\pi T}{{m_{h,n}}}}\ll1,
&&\epsilon^{3\rm D}_H\equiv\frac{6\lambda_H}{16\pi}\frac{T}{{m_{h,0}}}\ll1,
\\\no
&&\epsilon^{3\rm D}_S\equiv\frac{\lambda_SN}{16\pi}\frac{T}{m_{s,0}}\ll1,\\\no
&&|\epsilon^{3\rm D}_{HS}|\equiv\frac{2|\lambda_{HS}|\sqrt{N}}{16\pi}\frac{T}{\max[{m_{h,0}, m_{s,0}}]}\ll1,
\ea
where $m_{h,0}, \,m_{s,0}$ are the effective masses of the corresponding zero modes.~\footnote{In the case of interest $m_H^2(T)<0$ and the origin is unstable. Since our perturbative estimates only make sense around an energy minimum, Eq. \eqref{pertT} should be interpreted as relations involving the zero mode masses around the vacuum, where the Higgs mass is $m_{h,0}\sim\sqrt{2}|m_H(T)|$.} Comparing leading and subleading corrections to 4-point functions of $S,H$ we find additional constraints. For example, requiring that the $S$ and Higgs 1-loop contributions to $\langle S S S S\rangle$ be smaller than the tree-level respectively give ${\lambda_SNT}/({8\pi m_{s,0}})\ll1$ and ${\lambda_{HS}^2T}/({8\pi m_{h,0}})\ll\lambda_S$, which when combined imply
\ba\label{pertThS}
\frac{\lambda_{HS}^2{N}}{(8\pi)^2}\frac{T^2}{m_{h,0}m_{s,0}}\sim(\epsilon^{3\rm D}_{HS})^2\frac{\max[{m_{h,0}, m_{s,0}}]^2}{m_{h,0}m_{s,0}}\ll1.
\ea
This may be actually a stronger condition than the last one of \eqref{pertT} depending on the masses involved as well as the actual size of $\epsilon^{3\rm D}_{HS}$. We will see below that a precise determination of the expansion parameter associated to $\lambda_{HS}$ is fortunately not necessary because its perturbativity is always guaranteed by stability and $\epsilon_{H,S}^{3\rm D}\ll1$.

Using  eq.(\ref{zeromodemass1}) for the 3D effective masses we can finally estimate the 3D expansion parameters in analogy with the $\lambda\varphi^4$ we discussed above. Let us focus on $\epsilon_H^{3D}$ first.  By eq.~(\ref{strongPT}),  $|m_{h,0}|/ (\sqrt{2\lambda_H}T)\sim h(T)/T\gtrsim 1.2-1.5$, the first of eqs.~(\ref{pertT}) implies 
\beq 
\epsilon_H^{3D}\sim  {\frac{\sqrt{3\epsilon_H}}{4}}\times \frac{T}{h(T)}\lesssim 0.02-0.03
\eeq
where we used the value of $\lambda_H$ renormalized at the weak scale. At higher scales $\lambda_H$ decreases so that the effective 3D coupling gets more suppressed. By the above equation we conclude that $\epsilon_H^{3D}$ is rather small and hence that we do not expect the need to chase higher order effects associated to this coupling. We expect similar conclusions concerning the 3D fate of the electroweak gauge couplings.
Notice also that 
this is just an upper bound. There is in principle no limitation in making $h(T)/T$ bigger by taking $N$ larger. Indeed for $h(T)/T$ 
big enough, the gap in the Higgs (as well as $W,Z, t$) gets larger than $T$, so that there is no significant 3D dynamics associated to this mode, as we shall discuss in more detail at the end of this section.

Regarding $\epsilon_S^{3\rm D}$ we similarly have~\footnote{The relation coincides with  eq.(\ref{3vs4}) as it is independent on $N$.}
\ba\label{pertT1}
\epsilon_S^{3\rm D}=\frac{\sqrt{3}}{2}\sqrt{\epsilon_S}.
\ea
In this case however we are not forced to go to very small values of $\epsilon_S$. To the contrary, eq.~\eqref{Nrough} shows that the smallest $N$ correspond to largish $\epsilon_S> 0.01$, which in turn implies $\epsilon_S^{3\rm D}\gtrsim 0.1$. Under these circumstances, while the perturbative series in 4D may still converge well, thermal 3D perturbation theory may require re-summation in order to make more firm statements. Luckily this can be done at leading order in the $1/N$ expansion, as we shall describe in the next section.  

We should finally discuss perturbativity for $\epsilon_{HS}^{3D}$. By using the stability constraint $\lambda_{HS}^2\leq \lambda_H\lambda_S$ and eq.~(\ref{pertT}) we have
\beq
|\epsilon_{HS}^{3D}|\leq \sqrt{\frac{2}{3}}\sqrt {\epsilon_H^{3D}\epsilon_S^{3\rm D}}\sqrt\frac{\min[{m_{h,0}, m_{s,0}}]}{\max[m_{h,0}, m_{s,0}]}\leq \sqrt{\frac{2}{3}}\sqrt {\epsilon_H^{3D}\epsilon_S^{3\rm D}}
\eeq
somewhere in between the other two expansion parameters. Similarly, \eqref{pertThS} is automatically satisfied provided we combine stability and 3D perturbativity of $\lambda_{S,H}$.

Let us now discuss the impact of the sizable contribution to the thermal masses of $h,W,Z, t$ that a large $h/T$ could imply. In ordinary QFT, at small $N$ and at weak coupling, thermal corrections to masses are parametrically smaller than the scale $\sim \pi T$ controlling the Matsubara spectrum. This statement applies to all corrections, including those induced via the effect of thermal loops on the VEVs of scalars. Given for instance a trilinear coupling $g$, the condition for a small thermal correction to masses is roughly $ g^2 T^2\lesssim \pi^2 T^2$, which parametrically coincides with the smallness of the loop expansion parameter $g^2/8\pi^2$. Notice in particular that at weak couplings the zero modes remain lighter that $\sim \pi T$, even if the relative effect is large, given the lightness of these modes at tree level. The consequence on the  3D dynamics  are just as we discussed above. However, as our model clearly illustrates, the situation changes drastically when considering QFT at large-$N$, even at weak coupling. As we discussed, in the limit $N\to \infty$ with $\epsilon_S$ and $\epsilon_{HS}$ finite but perturbatively small, the tadpole correction of Fig. \ref{fig:tadpole} to the thermal Higgs mass scales like $\epsilon_{HS}\sqrt N\to \infty $ and can consistently become much larger than $\pi T$. When that is the case and $\lambda_{HS}<0$ the resulting VEV of $h$ will also shift  the masses of $W,Z, t$ 
  way above $\pi T$. More precisely that happens when $\epsilon_{HS}\sqrt N\gg 1$.
  
It is worth, to gain a better overview on the dynamics, to focus for a moment on the case $\epsilon_{HS}\sqrt N\gg 1$, even though 
this would require $N$ much larger than the lowest allowed value. In this case, the heaviest SM states $h, W,Z, t$ decouple from the thermal dynamics: their density and all their effects, including those on the effective potential, are Boltzmann suppressed.  In particular they decouple from the  3D 
stochastic  dynamics.
 Notice, in contrast, that in the $N$ scaling we are considering $\epsilon_S$ is fixed: as long as perturbation theory applies, the mass $m_{s,0}$ of the zero mode of $S$, see eq.~(\ref{zeromodemass1}),  remains $\ll \pi T$, and this mode dominates the thermal dynamics. When $\epsilon_{HS}\sqrt N\gg 1$, with $\epsilon_S,|\epsilon_{HS}|\ll 1$ the thermal spectrum has then  the rough structure
\beq
m_h^2,m_W^2,m_Z^2, m_t^2\gtrsim \pi^2 |\epsilon_{HS}|\sqrt N T^2 \gg \pi^2 T^2 \gg \pi^2 \epsilon_S T^2 \sim m_s^2\, .
\eeq
In order to systematically compute physical quantities one should then use an effective field theory approach in three steps: 1) integrate out the heavy SM states $h,W,Z,t$ matching to an EFT for the remaining light (and nearly decoupled)  SM states plus  $S$; 2) run this EFT down to the compactification scale $\sim \pi T$; 3) match to a 3D EFT for $S$.
Notice that the dynamics that determines the thermal masses of  $h, W,Z, t$ takes place at much lower energies, at or below $T$.
In a sense the situation is similar to the case, naturally realized in supersymmetry, where  a flat direction field controlling the mass of heavy states is stabilized by some  IR dynamics.  The difference with respect to our  case is
that  here the operator $S^2$ controlling the heavy masses is a genuine composite, in the sense that $\langle S\rangle =0$. Indeed the large size of $\langle S^2\rangle$ is not driven by the flatness of the potential but by the coherent pile-up of thermal fluctuations from $N$ fields. We can however formulate the dynamics  in a way that is not too different from the case of a flat direction. It just suffices to introduce an auxiliary field $\sigma$ mediating the  $(S^2)^2$ interaction, as it is standard in the study of large-$N$ models like the Gross-Neveu model or the $O(N)$ model.  As discussed in the next section,  one can start from the equivalent Lagrangian eq.~(\ref{aux}), which reduces to the original one by first integrating out $\sigma$.  However, according to the EFT
picture outlined above, it is more convenient to first integrate out $h$ as well as $W,Z,t$, working around a background with non-vanishing $\sigma$. In such a way one derives an EFT redundantly written in terms of $S$ and $\sigma$, where  all degrees of freedom are light. This effective Lagrangian is exactly quadratic in $S$, and can be studied in a $1/N$ expansion by first integrating out all of $S$. We will partly illustrate that in the next section. 

The extreme limit $\epsilon_{HS}\sqrt N\gg 1$ just described is formally interesting. However, the estimates in \eqref{Nrough} reveal that typical values of $\epsilon_{HS}\sqrt N$ are usually far away from it. In these more realistic cases there is no big separation of scales, so we do not need to construct the effective theory as outlined in the previous paragraph. As we will see in the analysis of the next subsection, however, it is still useful to integrate-in the field $\sigma$ (which controls the mass of $h, W,Z, t$) and keep all non-linear terms in this quantity; on the other hand, it suffices to work at one loop order in the SM couplings.

\subsection{The effective potential}
\label{sec:effV}

In this subsection we will find a more accurate approximation for the effective potential and to the conditions necessary to satisfy \eqref{strongPT}. 

We begin by recalling that eq.~(\ref{Nrough}) indicates that the region of parameter space where sphalerons are switched off at high $T$ lies always at large $N$. Moreover the value of $N$ is minimized for the largest $\epsilon_S$,  compatibly with
perturbativity.  These two joint facts prompt us to use the standard large-$N$ methodology to re-sum all orders in $\epsilon_S$. Indeed, neglecting for the moment the effects of $\lambda_{HS}$, the loop expansion in the $S$ sector is easily seen to correspond to a series in $\lambda_S^p N^q\propto \epsilon_S^pN^{q-p}$ with $p\geq q$. Treating $\epsilon_S$ as a fixed and not necessarily small parameter, the large-$N$ resummation consistently captures at leading order  all the terms with $p=q$, with the next-to-leading terms ($p=q+1$) formally suppressed by $O(1/N)$. See the right of Fig. \ref{fig:tadpole} for a sample of leading diagrams.

In the purely 4D case ($T=0$), while the resummation of the leading $p=q$ series makes perfect sense, the dynamical regime
$\epsilon_S\gtrsim O(1)$, where the effects would be most dramatic and interesting, lies unfortunately in the UV, where the theory
is out of control.~\footnote{The situation is famously reversed in Gross-Neveu model in 2D, where the coupling is asymptotically free and the strong regime lies controllably in the IR.} However, at finite temperature, as we already reviewed, the effective coupling becomes stronger in the 3D regime. In this situation, the resummation of the leading series in $1/N$ can capture 
more dramatic effects. For example, interpreting the mass $m_{s,0}$ of the zero mode of $S$ as a free parameter, say by suitably tweaking the bare squared mass of $S$ (in particular allowing it to be negative), the IR dynamics in 3D potentially displays various phases that are all tractable at large-$N$. In the strict $m_{s, 0}=0$ limit the system flows to a CFT (the so-called $O(N)$ model) in the far IR. The typical scale of the flow is $\sim \lambda_SNT/8\pi$, where the coupling $\lambda$ becomes strong but the $1/N$ approximation remains reliable. For 
    $0<m_{s, 0}\lesssim \lambda_SNT/8\pi $ we have a strongly coupled gapped phase, while for $m_{s, 0}\gtrsim O(\lambda_SNT/8\pi)$ we  get back to a weakly coupled regime, where however the effective coupling $\epsilon_S^{3D}\sim\lambda_SNT/(8\pi m_{s, 0})$ is bigger than $\epsilon_S$; in particular, in the generic high $T$ phase we have seen that $\epsilon_S^{3D}\sim \sqrt{\epsilon_S}$.  
In our model \eqref{mod} we are always in this latter case and therefore it is the less dramatic, but still quantitatively important, series in $\epsilon_S^{3D}$ that our large-$N$ resummation shall capture.
To be more quantitative, let us estimate the range of $\epsilon_S$ compatible with the perturbative definition of our model.  For that purpose, observe that at leading order in $1/N$, and considering the realistic limit $\epsilon_H\ll\epsilon_S$, the beta function $\mu d\epsilon_S/d\mu=2\epsilon_S^2$ develops a Landau pole at $\Lambda_L=\mu\exp[1/(2\epsilon_S(\mu))]$. Having a consistent theory below $\Lambda=100$ TeV $<\Lambda_L$ then translates into the perturbativity requirement $\epsilon_S(m_t)\lesssim0.07$. We confirmed numerically that the inclusion of all other couplings does not alter this upper bound significantly. In the following we will therefore stick to the domain $\epsilon_S(m_t)\lesssim0.07$.  Such a value corresponds to $ \sqrt{\epsilon_S}\sim 0.25$
for which resummation of the leading 3D effects can make an almost $O(1)$ effect.

Having identified a reliable and systematic re-summation suitable for our model, let us now go back to our original task: finding the effective potential. An important simplification in this analysis comes from the fact that the vacuum of $S$ is trivial at all temperatures whenever \eqref{muS} holds. We have already shown this in the simplified analysis of Section \eqref{firstapprox}, but it turns out it is in fact a general result, as argued in Appendix~\ref{app:s=0}.

Now, the standard technique to resum the leading loop effects at large-$N$ in the $O(N)$ model involves the introduction of an auxiliary field $\sigma$ mediating the interactions of interest. In practice we shall add a  trivial term $\frac{1}{4\lambda_S}\left(\sigma-\lambda_SS^2-2{\lambda_{HS}}H^\dagger H\right)^2$ to eq.~(\ref{mod}),  such that the part of the Lagrangian that depends only on the scalar fields becomes:
\ba\label{aux}
{\cal L}&=&D_\mu H^\dagger D^\mu H-\left({m_H^2}+\frac{\lambda_{HS}}{{\lambda_S}}\sigma\right)H^\dagger H-{\lambda_H}\left(1-\frac{\lambda_{HS}^2}{\lambda_H\lambda_S}\right)(H^\dagger H)^2\\\no
&+&\frac{1}{2}\partial_\mu S\partial^\mu S-\frac{1}{2}(m_S^2+\sigma)S^2+\frac{1}{4\lambda_S}\sigma^2.
\ea
Of course when integrating out $\sigma$ this trivially gives back our Lagrangian. However, in order to organize perturbation theory as an expansion in $1/N$, it is convenient to first integrate out $S$. The key property of eq.~\eqref{aux} is that $S$ appears quadratically, so that it can be integrated out exactly to give:
\ba\label{aux1}
{\cal L}\to{\cal L}_{\rm eff}&=&D_\mu H^\dagger D^\mu H-\left({m_H^2}+\frac{\lambda_{HS}}{{\lambda_S}}\sigma\right)H^\dagger H-{\lambda_H}\left(1-\frac{\lambda_{HS}^2}{\lambda_H\lambda_S}\right)(H^\dagger H)^2\\\no
&+&\frac{1}{4\lambda_S}\sigma^2+N\Gamma[m_S^2+\sigma, \partial],
\ea
where $\Gamma[m_S^2+\sigma, \partial]$ is a calculable function of $\sigma$ and its derivatives (Note we also used the fact that $\langle S\rangle=0$ in deriving \eqref{aux1}. We will assume this is understood in the following.). The most salient property of the above result is that in the limit $N\to \infty$ with $\lambda_SN$ fixed, the action for $\sigma$ in the second line is formally of order $N$. Neglecting for the moment the first line involving the interaction with $H$, this implies that $1/N$ is the loop expansion parameter for the self-interactions of $\sigma$. This is easily seen by going to a canonical basis $\sigma \to \sqrt{\lambda_S}\sigma\sim \sigma/\sqrt N$ and expanding $\Gamma$ in powers of $\sigma$: the $n$-point couplings scale like 
$(1/\sqrt N)^{n-2}$ as befits $1/N$ being the loop counting parameter. By the same rescaling, the  $\sigma H^\dagger H$ coupling becomes $\lambda_{HS}/\sqrt{\lambda_S}$, which by the stability condition is $\leq \sqrt{\lambda_H}$. We thus conclude that, in the large-$N$ limit with the SM couplings and $\epsilon_{S}$ and $\epsilon_{HS}$ fixed, $\sigma$ interactions come in two classes: self interactions, with strength controlled by powers of $1/N$, and interactions with the Higgs, whose strength is
 comparable, at most, to that of the SM couplings. 

Eq.~\eqref{aux1} contains all the quantum fluctuations from $S$  at leading order in $1/N$,
with subleading effects associated to $\sigma$ fluctuations  not yet computed. When treating the couplings as the bare ones, the above Lagrangian should then not involve any UV divergence associate to the leading $S$ loops. This is easily checked. 
$\Gamma$ involves indeed a logarithmic UV divergent $\sigma^2$ piece that matches precisely the UV divergence in the tree coefficient $1/4\lambda_S$. Moreover,  using the beta functions given in \eqref{betafunctions}, one can easily check
that the other two combinations of couplings $\lambda_{HS}/\lambda_{S}$ and $\lambda_H-\lambda_{HS}^2/\lambda_S$ are RG invariant when considering the contributions that are purely induced by $S$-loops and  leading in the $1/N$ expansion. It follows that the corresponding bare couplings are free of the associated UV divergences, as it should. This result means for instance that  the two point function of $\sigma$ from eq.~\eqref{aux1} will be determined in the leading log approximation by  $\lambda_S$ renormalized at the largest scale
among $T$, $\sqrt {\langle \sigma \rangle}$ and the external momentum $\partial$. In practice when considering thermal loops 
the relevant scale will be $T$. Notice however that $\Gamma$ also contains finite corrections coming from $3D$ physics, whose relative size is controlled by $\sqrt{\epsilon_S}$ and which we %could in principle 
will take into account.

With all the above comments in place we can now compute  the effective potential. Motivated by the estimates in section \ref{firstapprox} a reasonable approximation would be to work at leading order in the $1/N$ expansion (but all orders in $\epsilon_S$) and at 1-loop in the SM couplings. Proceeding in the standard way we decompose the fields in their classical background plus their fluctuations. Ignoring the eaten Goldstones, which we will treat separately, in the scalar sector we have
\beq
\sigma= \sigma_c+\delta \sigma\qquad\qquad H=\left(\begin{matrix}
0\\
\frac{h+\delta h}{\sqrt{2}}\end{matrix}\right).
\eeq
When expanding the action at quadratic order  in $\delta \sigma$ and $\delta h$ we find a slight complication from 
a mixing term  arising from the cubic interaction $\sigma H^\dagger H$. It is then convenient to integrate over the quantum fluctuations in two steps:  we first perform a suitable shift of $\delta \sigma$ to eliminate the mixing and then compute the fluctuation determinant for the resulting diagonal quadratic action. The first step corresponds to 
\beq
\delta \sigma \to \delta \sigma +\frac{\lambda_{HS}}{\lambda_S}G_{\sigma\sigma} h\delta h
\eeq
with
\beq
G_{\sigma\sigma}= \left (\frac{1}{2\lambda_S}+\frac{N\delta^2 \Gamma}{\delta \sigma\delta\sigma}\right )^{-1}
\eeq
 the $\delta \sigma$ propagator around the background. The $\delta h$ self-energy gets corrected according to
 \ba\label{hhdressed}
 \Sigma_{hh}&\equiv&\partial^2+m_H^2+\frac{\lambda_{HS}}{\lambda_S}\sigma_c+{3}\lambda_H\left(1-\frac{\lambda_{HS}^2}{\lambda_H\lambda_S}\right)h^2\\\no
 &\to& \Sigma_{hh}+\frac{\lambda_{HS}^2}{\lambda_S^2}G_{\sigma\sigma} h^2.
 \ea
After this diagonalization we notice that the $\delta\sigma$ fluctuation determinant contributes a term that purely depends on $\sigma_c$ and which is $1/N$ suppressed with respect to the leading action for $\sigma$. In accordance to our goals we neglect this term. The $\delta h$ contribution, determined by the ${\mathrm {Tr}}\, {\ln}$ of the self energy operator in eq.~\eqref{hhdressed}, corresponds to the SM Higgs contribution to the 1-loop effective potential, dressed by an infinite class of $S$ loops, which resum the leading series in $\epsilon_S$ in the large-$N$ limit. This computation is made involved by the non trivial momentum dependence of $G_{\sigma\sigma}$. In view of that, we have chosen to simplify our computation by making the approximation $G_{\sigma\sigma}\simeq 2\lambda_S(T)$. The error this entails corresponds to $O(\sqrt{\epsilon_S})$ relative to the $h$-loop contribution to the effective potential. In view of the smallness of the leading Higgs loop contribution  compared to the combined effects of $W,Z, t$, we  estimate this is a fair approximation. In principle, with some extra effort, this approximation could be dropped. We plan to reconsider this in the future, though it seems rather clear this is not going to change our estimates on the range of parameters by more than a few percent.

With all the above comments in place, in particular as concerns our approximations, after some straightforward algebra, at last, the effective potential for $h,\sigma_c$ reads
\ba\label{VeffT1}
V_{\rm eff}(h,\sigma_c)&=&\frac{1}{2}\left({m_H^2}+\frac{\lambda_{HS}}{{\lambda_S}}\sigma_c\right)h^2+\frac{\lambda_H}{4}\left(1-\frac{\lambda_{HS}^2}{\lambda_H\lambda_S}\right)h^4\\\no
&-&\frac{1}{4\lambda_S}\sigma_c^2+NV_0(m_S^2+\sigma_c)\\\no
&+&V_{0}\left(m_H^2+\frac{\lambda_{HS}}{\lambda_S}\sigma_c+\lambda_H\left(3-\frac{\lambda_{HS}^2}{\lambda_H\lambda_S}\right)h^2\right)\\\no
&+&3V_{0}\left(m_H^2+\frac{\lambda_{HS}}{\lambda_S}\sigma_c+\lambda_H\left(1-\frac{\lambda_{HS}^2}{\lambda_H\lambda_S}\right)h^2\right)\\\no
&+&6V_1\left(\frac{g^2}{4}h^2\right)+3V_1\left(\frac{g^2+g'^2}{4}h^2\right)+12V_{1/2}\left(\frac{y_t^2}{2}h^2\right)
%\\\no
%&+&{\cal O}(1/N,\sqrt{\epsilon_S}\epsilon_{h, 3D},\epsilon_{\rm SM, 3D}^2),
\ea
where all the couplings should be taken renormalized at $\mu \sim T$ and the functions $V_j$ are defined in Appendix~\ref{app:V}. The contributions in the above equation are easily identified. Besides the ``tree" level terms in the first two lines,
the third line corresponds to the physical Higgs scalar fluctuation $\delta h$, the fourth to the eaten Goldstones in Landau ($\xi=0$) gauge, the fifth to respectively $W,Z$ and $t$. Notice that the difference in the argument of the function $V_0$ for Higgs and Goldstones is also determined by the mixing with $\sigma$ we mentioned above.

Overall, \eqref{VeffT1} includes all 1-loop effects and re-sums {\emph{all powers}} of $\epsilon_S$ at leading $1/N$ order, but ignores genuinely 2-loop contributions involving the SM couplings --- since expected to be smaller than those solely controlled by $\epsilon_S$ --- as well as corrections of order $\sqrt{\epsilon_S}$ on the already small Higgs contribution in the third line. All effects scaling as $\lambda_{HS}N$, which play a crucial role in symmetry non-restoration, come from $\sigma_c\lambda_{HS}/\lambda_S$ and have thus been taken into account in \eqref{VeffT1}. An example of re-summed contributions is shown in the right panel of Fig. \ref{fig:tadpole}.

\begin{figure}%[t]
\begin{center}
\includegraphics[width=8.4cm]{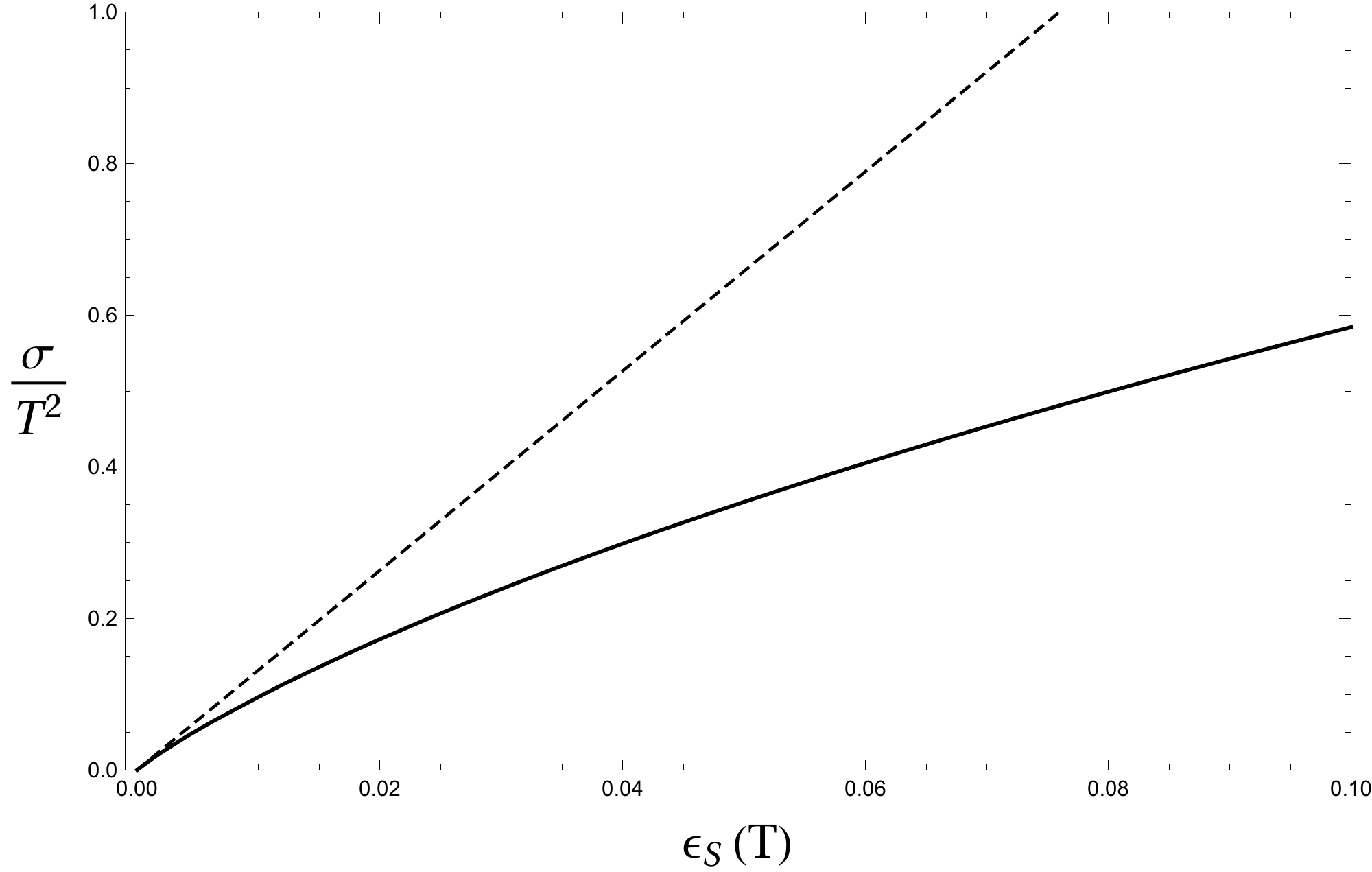}
\caption{The solid curve shows the exact classical solution $\sigma(s^2=0)/T^2$ of \eqref{aux1} with $H\to0$, as a function of $\epsilon_S(\mu=T)$. The dotted line is the approximate expression $\sigma/T^2={\lambda_SN}/{12}$ obtained at leading order in an expansion in $\epsilon_S\ll1$.
}\label{figsigma}

\bigskip

\bigskip
%\vspace{0.4cm}

\includegraphics[width=8.5cm]{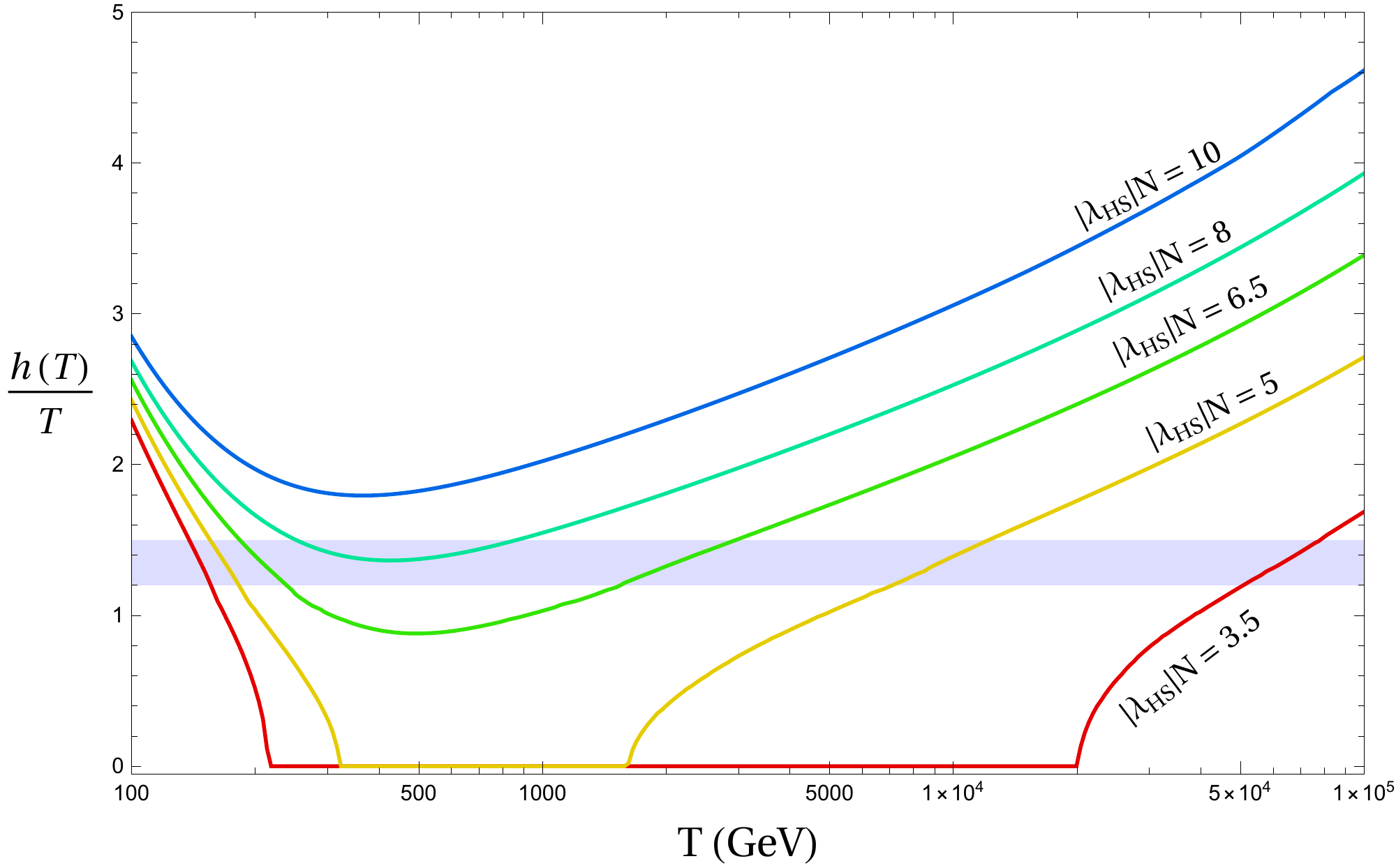}
\caption{Here we show $h/T$ as a function of the temperature for various values of $\lambda_{HS} N$ and $m_s=0, \epsilon_S=0.03, N=1000$. The filled area identifies $1.2\leq h/T\leq1.5$. As discussed, these curves have a minimum at around $100 - 1000$ GeV, where the transition between the high-$T$ and low-$T$ behavior takes place. The growth at high temperatures is due to the running of $\lambda_H$. 
}\label{fighoverT}
\end{center}
\end{figure}

The true effective potential for $h$ is found solving numerically the classical equations of motion for the constant configuration $\sigma_c(h^2)$ and replacing it back in \eqref{VeffT1}, i.e. 
\ba\label{36}
V_{\rm eff}(h)=V_{\rm eff}(h,\sigma_c(h^2)). 
\ea
Yet, before presenting our numerical results it is instructive to investigate the ideal limit in which only the second line of \eqref{VeffT1} is kept. This limit is useful to quantify for which values of $\epsilon_S$ our model is expected to be comfortably under perturbative control at finite $T$. The numerical solution $\sigma_c$ in this simplified case, and for the high $T$ regime $\sigma_c\gg m_S^2$, is shown in Fig. \ref{figsigma} as a function of $\epsilon_S(\mu=T)$. The exact result (solid line) approaches the expression $\sigma_c={\lambda_SN}T^2/{12}$ (dotted line) found at leading order in an expansion in $\epsilon_S$ (i.e. the one we used for example in \eqref{zeromodemass1}) only for $\sigma_c/T^2\ll1$. Quantitatively, the leading order expression $\sigma_c={\lambda_SN}T^2/{12}$ is larger by a factor $(1-2\sqrt{3\epsilon_S}+{\cal O}(\epsilon_S))$, that is $30-40\%$ already at $\epsilon_S\sim0.01$ and gives us a measure of the perturbative domain at high $T$. This result is especially important to us because physically $\sigma_c$ represents the thermal mass squared of the singlet (see also \eqref{0V}), whereas $\sigma_c\lambda_{HS}/\lambda_S$ is the dominant thermal correction to the Higgs mass squared, see \eqref{VeffT1}. What Fig. \ref{figsigma} demonstrates is that when $\epsilon_S\gtrsim0.01$ a numerical study of \eqref{VeffT1}, including in particular the full function $NV_0(m_S^2+\sigma_c)$, is necessary to obtain a careful assessment of the effective potential at finite $T$.  

%%%%%%%%%%%%%%%%%%
%%%%%%%%%%%%%%%%%%
%%%%%%%%%%%%%%%%%%
%%%%%%%%%%%%%%%%%%

We are finally ready to discuss our results. We derived $V_{\rm eff}(h)$ numerically according to \eqref{36} for a set of input values $\lambda_S(m_t), \lambda_{HS}(m_t), N$ and
\ba\label{masss}
m_s^2=m_S^2+\lambda_{HS}v^2.
\ea
We took $y_t(m_t)=0.934-0.951$, $g(m_t)=0.648$, $g'(m_t)=0.359$, $g_s(m_t)=1.17$, $\lambda_H(m_t)=0.128$ and $m_t=173~{\rm GeV}$, $m_h=125$ GeV. Consistently with our approximations in the effective potential, the RG evolution of the couplings is evaluated at 1-loop. The renormalization scale was fixed at the value $\mu=\sqrt{m_t^2+T^2}$ to minimize the logs in $V_j$ at large $T$ and avoid singularities at $T=0$. 

%%%%%%%%%%%%%%%%%%
%%%%%%%%%%%%%%%%%%

%\thispagestyle{empty}
\begin{figure}[tbp]
\begin{center}
\includegraphics[width=9.1cm]{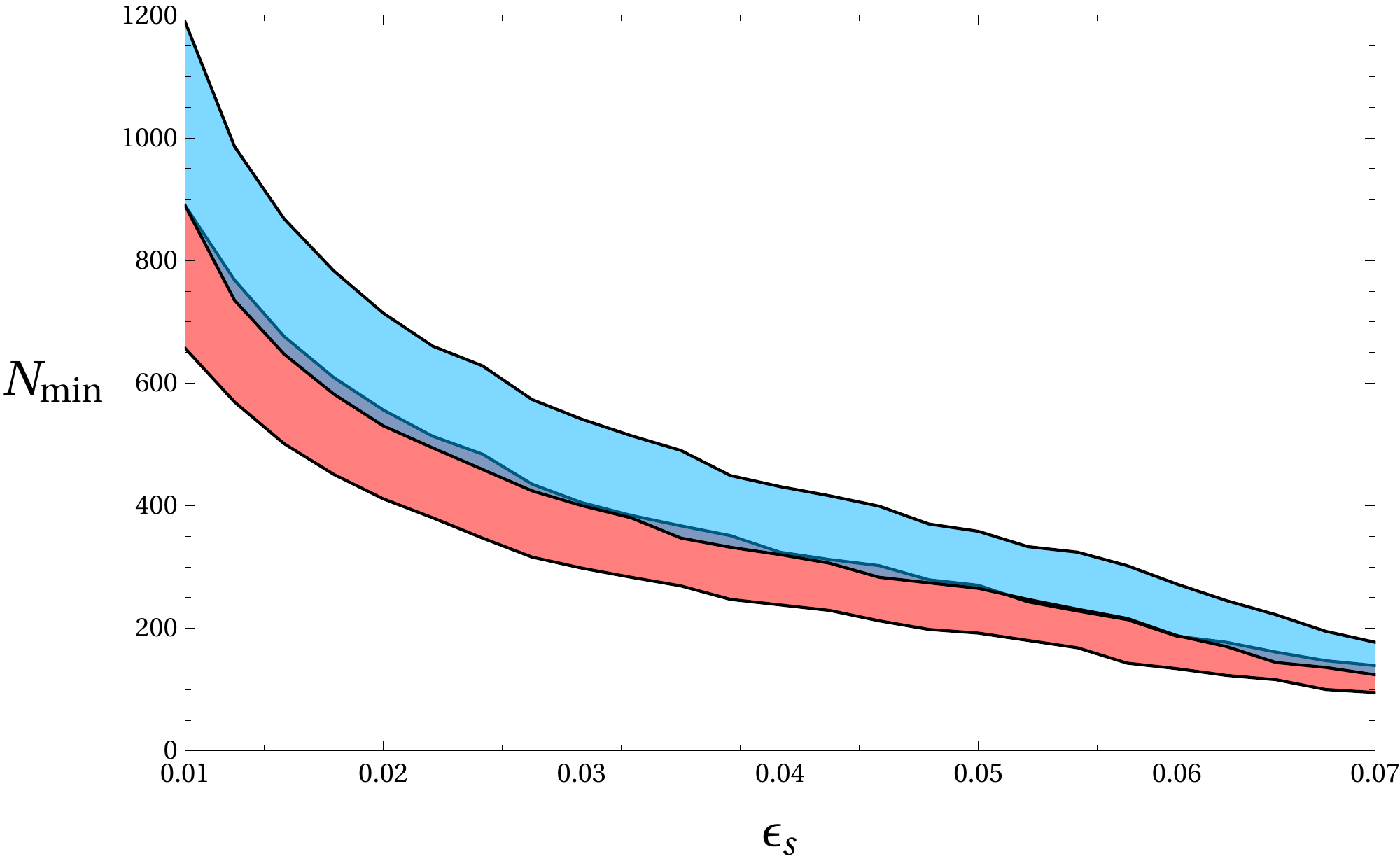}%\\\includegraphics[width=10cm]{Figs/epscontour.pdf}
\caption{The colored area shows the minimum value of $N$ for a given $\epsilon_S$ that gives $1.2\leq h/T\leq1.5$ for all $T\leq100$ TeV compatibly with the stability requirement $\lambda_{HS}^2\leq\lambda_H\lambda_S$. Here we took $m_s=0$ and $y_t(m_t)=0.934$ (red) $0.951$ (blue). %As discussed in the text the value $y_t(m_t)=0.934$ appears to provide a more realistic estimate of the actual bound. 
}\label{figNmin}

\bigskip

\bigskip

\includegraphics[width=9cm]{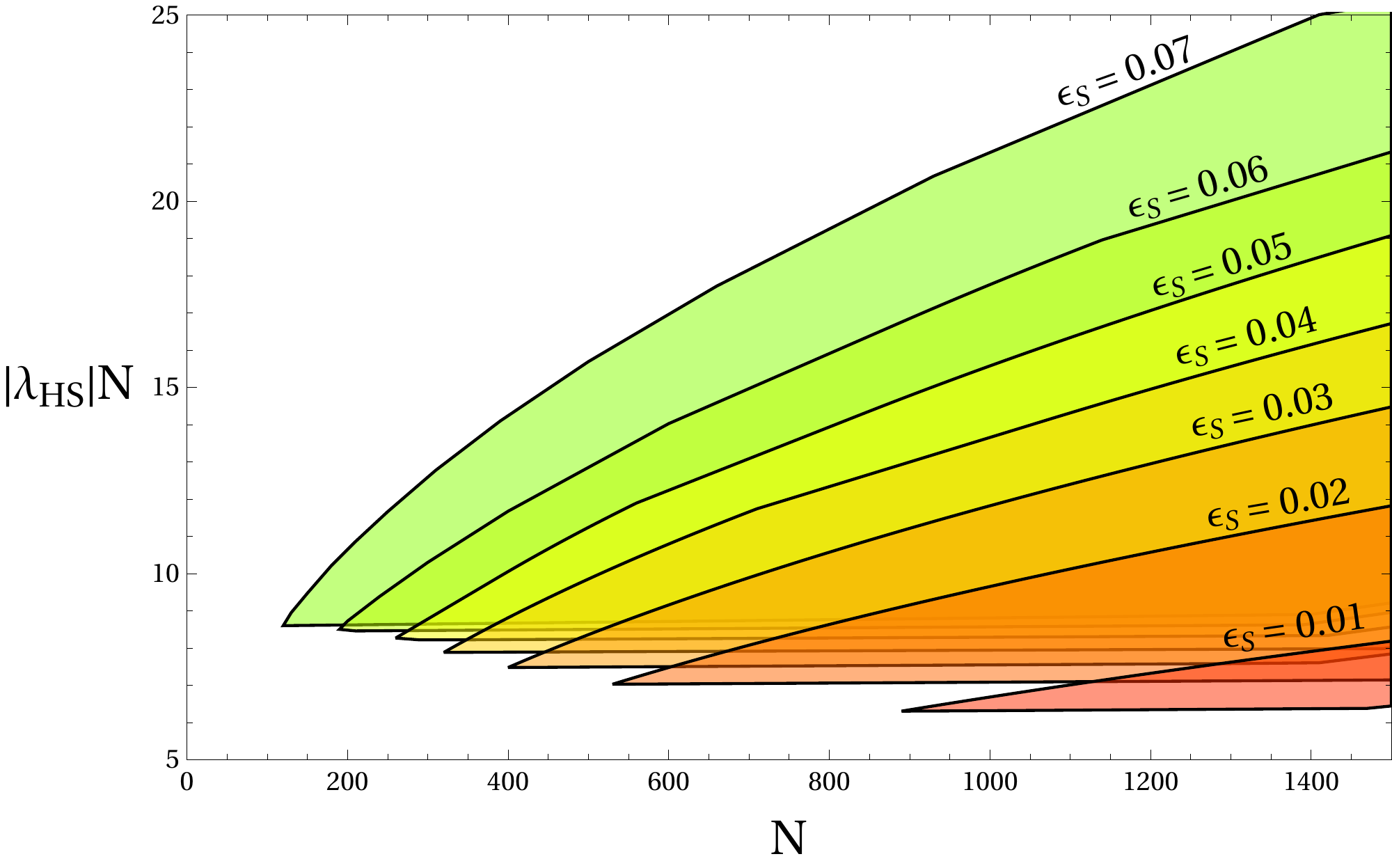}
\caption{Region with $h/T\geq1.2$ in the $N$, $\lambda_{HS}N<0$ plane. Here $m_s=0$, $y_t(m_t)=0.951$.
}\label{figMix}

\bigskip

\bigskip

\includegraphics[width=9cm]{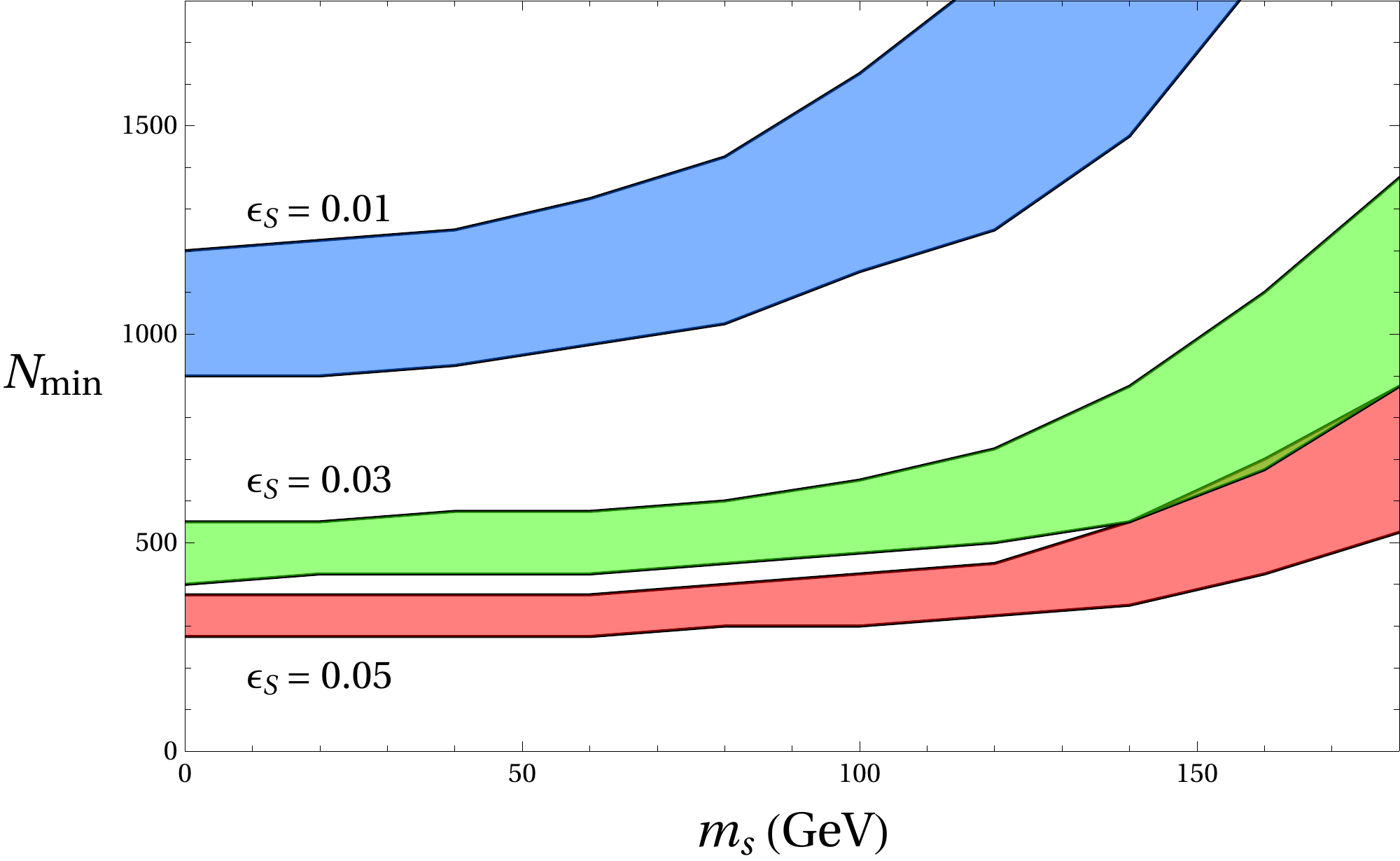}
\caption{$N_{\rm min}$ as a function of $m_s$ for a few representative values of $\epsilon_S$ (here $y_t(m_t)=0.951$). In the colored region $1.2\leq h/T\leq1.5$ and the model is stable for all $T\leq100$ TeV.
}\label{figMass}

\end{center}
\end{figure}

%%%%%%%%%%%%%%%%%%
%%%%%%%%%%%%%%%%%%

First we note that, as already anticipated when discussing \eqref{Nrough}, the function $h(T)/T$ increases at high $T$ due to the decrease in $\lambda_H(\mu)$ and has a minimum at around the temperature at which the transition between high and low $T$ occurs. This implies that the requirement that sphalerons are switched off at all $T<\Lambda=100$ TeV is dominated by low temperatures, which motivates a posteriori the approximation made in \eqref{strongPT}. This behavior is shown in fig. \ref{fighoverT}. Our analysis also confirms the result \eqref{Nrough} quantitatively: $h\gtrsim T$ requires large $N$ and a sizable $\epsilon_S$. Specifically, $N\gtrsim100-200$ seems unavoidable if we require the description to stay perturbative up to $\Lambda=100$ TeV.

In fig. \ref{figNmin} we show, for $\epsilon_S\in[0.01,0.07]$ and $m_s=0$, the value $N_{\rm min}$ corresponding to the smallest $N$ that gives $h/T\geq1.2-1.5$ compatibly with the stability condition $\lambda_{HS}^2<\lambda_H\lambda_S$ all the way up to $T=100$ TeV. As already evident from \eqref{Nrough}, $N_{\rm min}$ is obtained when maximizing $\epsilon_S$. Moreover, for a given $\epsilon_S$ the minimum $N$ is found at the maximally allowed $|\lambda_{HS}|$. The ratio $\lambda_{HS}^2/\lambda_H\lambda_S$ increases with the RG scale, so that that the stability constraint is dominated by large RG scales: demanding stability up to our UV cutoff translates into $|\lambda_{HS}(m_t)|/\sqrt{\lambda_S(m_t)\lambda_H(m_t)}\lesssim0.4-0.6$. In this respect it is important to stress that the 1-loop approximation of the RG overestimates the drop in $\lambda_H(\mu\gg m_t)/\lambda_H(m_t)$ when $y_t(m_t)=0.951$ is used, and hence over-constraints $|\lambda_{HS}|$, whereas an evolution closer to a 3-loop analysis is obtained for the lower value $y_t(m_t)=0.934$. In this sense the choice $y_t(m_t)=0.934$ appears to be more accurate. As the figure shows, the weaker drop in $\lambda_H$ results in a weaker bound on $N$. Finally, we verified that for $N\gg N_{\rm min}$ and $\epsilon_S\to0$ our numerical solution $h(T)$ approaches \eqref{honTrough}, as it should.

Additional information is provided by fig. \ref{figMix}. Taking the conservative value $y_t(m_t)=0.951$, and again $m_s=0$, the plot shows what region in the $N,\lambda_{HS}N<0$ plane is selected by requiring $h/T\geq1.2$ and stability up to the UV cutoff, for a few choices of $\epsilon_S$. Note that $|\lambda_{HS}|N\gtrsim6-9$, compatibly with what we found in Section \ref{firstapprox}, where the smaller values are obtained for smaller $\epsilon_S$.

Another relevant question we can address with \eqref{VeffT1} concerns the allowed zero-temperature mass \eqref{masss} of the scalar. Crucially, $S$ cannot be much heavier than $m_t$ because otherwise it would not be able to generate large enough thermal corrections to the Higgs mass to win over the positive top quark contribution. A more quantitative assessment of this expectation is given in Fig. \ref{figMass}, where we present the minimal value of $N$ required to suppress sphaleron processes at all temperatures as a function of $m_s$, and for a few representative values of $\epsilon_S$.

It is worth emphasizing that, while we have argued that $N\gtrsim100-200$ is unavoidable within the minimal model \eqref{mod} (see however possible extensions discussed in Section~\ref{sub:smallN}), $N$ cannot be arbitrary large. The reason is that in scenarios of EW baryogenesis one expects~\cite{Cohen:1990py}
\ba\label{etab}
\eta_b\sim c\sin\theta_{\rm CP}(20\alpha_w^5)/g_*, 
\ea
where $20\alpha^5_w$ measures the strength of the non-perturbative sphaleron effects (as in section \ref{sec:sphaleron}), $c$ is a function of the wall dynamics, $\theta_{\rm CP}$ a CP-violating phase, and $g_*$ the total number of relativistic degrees of freedom. In our scenario \eqref{etab} is generated by the UV dynamics at $T\sim\Lambda$ and having $\sin\theta_{\rm CP}\sim1$ is not a problem. However, with $g_*\sim N\gg100$ the asymmetry would be suppressed by $\sim g_{*,\rm SM}/g_*\sim 100/N$ compared to standard scenarios of EW baryogenesis. For the typical values considered here, $N\sim10^2-10^3$, the net effect is a mild ${\cal O}(1-10)$ suppression, but much larger values of $N$ are clearly not viable. Specifically, from \eqref{etab} we find the observed baryon asymmetry $\eta_b\sim10^{-10}$ can only be reproduced provided $N/c\lesssim 10^4$.

%%%%%%%%%%%%%%%%%%%
%%%%%%%%%%%%%%%%%%%
%\begin{figure}[t]
%\begin{center}
%\includegraphics[width=10cm]{Figs/Nminvsmass.pdf}
%\caption{Value of $N_{\rm min}$ as a function of $m_s$ and for a few representative values of $\epsilon_S$ ($y_t(m_t)=0.951$). Again, in the colored region $1.2\leq h/T\leq1.5$ and the model is stable for all $T\leq100$ TeV.
%}\label{figMass}
%\end{center}
%\end{figure}
%%%%%%%%%%%%%%%%%%%
%%%%%%%%%%%%%%%%%%%

\section{Phenomenology}
\label{sec:pheno}

In the previous section we have seen that $S$ has to have a mass around the weak scale. This opens the possibility of observing direct and indirect signatures of the new singlet, as discussed in the present section.

\subsection{Collider constraints}
\label{sec:coll}

We have already emphasized that $S$ does not acquire a VEV because of our assumption \eqref{muS}: the $O(N)$ symmetry remains unbroken. $S$ must thus be produced in pairs in accelerators, with rates proportional to $\lambda_{HS}^2N$, and show up as missing energy. The most dramatic effect consists in the generation of an invisible Higgs branching ratio, and is kinematically allowed only for $m_s^2<m^2_h/4$: 
\ba
{\rm BR}(h\to{\rm inv})=\frac{\Gamma(h\to SS)}{\Gamma_{\rm tot}}=\frac{\lambda_{HS}^2v^2N}{8\pi m_h\Gamma_{\rm tot}}\left(1-4\frac{m_s^2}{m_h^2}\right)^{1/2}.
\ea
Requiring ${\rm BR}(h\to{\rm inv})<10\%$ and conservatively assuming $|\lambda_{HS}|N=10$ (see Fig. \ref{figMix}) we get the severe bound $N\gtrsim10^6$. To avoid it in the following we will stick to the regime $m_s>m_h/2$. 

Direct production via vector boson fusion or associated production has been considered in~\cite{Curtin:2014jma}. According to that study, for $m_h/2<m_s<{\cal O}(m_t)$ there exist realistic chances of discovering $S$ via direct production at a future $100$ TeV pp collider as long as $\lambda_{HS}\sqrt{N}\gtrsim0.2-0.4$, which for $|\lambda_{HS}|N=10$ reads $N\lesssim{\cal O}(10^3)$.

Besides direct processes, virtual effects can be important. Because the singlet does not mix with $H$, all couplings of the Higgs boson to the SM vector bosons and fermions are standard at tree level. Yet, 1-loop diagrams with virtual $S$'s modify them offering an interesting indirect probe of the model. The most important effect in the $N\to\infty$ limit concerns the Higgs wave function and corresponds to a universal reduction by a factor $(1-\delta Z_H/2)$ of on-shell Higgs couplings with respect to their SM value. One finds
\ba
\delta Z_H&=&\frac{\lambda_{HS}^2Nv^2}{8\pi^2m_s^2}\int_0^1 dx\frac{x(1-x)}{1-\frac{m_h^2}{m_s^2}x(1-x)}\\\no
&=&\frac{\lambda_{HS}^2Nv^2}{48\pi^2m_s^2}f(m_h^2/m_s^2),
\ea
where $f(m_h^2/m_s^2)$ is a decreasing function of $m_s$, of order $f=3.5 (1.5)$ for $m_s=70 (100)$ GeV and quickly approaching $f=1$ for larger masses. For $m_s={\cal O}(m_h)$, and using $|\lambda_{HS}|N=10$ as a benchmark value, $\delta Z_H\sim({\lambda_{HS}^2Nv^2})/({48\pi^2m_s^2})\sim1/N$. Future linear colliders might be able to test $N\sim100-1000$ \cite{Dawson:2013bba} and thus a sizable portion of the relevant parameter space.

%%%%%%%%%%%%%%%%%%
%%%%%%%%%%%%%%%%%%
\begin{figure}[t]
\begin{center}
\includegraphics[width=5cm]{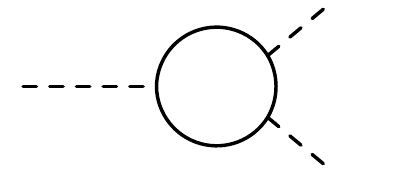}~~~~~~~~~~\includegraphics[width=5cm]{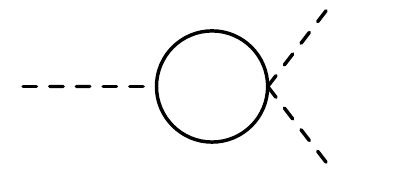}
\caption{The two 1-loop diagrams contributing to the Higgs trilinear.
}\label{tril}
\end{center}
\end{figure}
%%%%%%%%%%%%%%%%%%
%%%%%%%%%%%%%%%%%%

Other corrections are less important. Consider for instance the corrections to the Higgs trilinear arising at 1-loop from  the diagrams in Figure~\ref{tril}. As $S$ is light, the result can only  roughly be described as a momentum independent correction to the  trilinear.  We are here nonetheless only interested in a rough estimate of the size of the effects. The triangle diagram in the left of Fig.~\ref{tril} gives a correction  $\sim{\lambda_{HS}^3Nv^2}/({24\pi^2\lambda_Hm_s^2})$ relative  to the SM trilinear. By taking $|\lambda_{HS}|N=10$ this becomes $\sim(100/N)\delta Z_H\sim 100/N^2$. This is only comparable to the wave function effect in the limiting case $N\sim 100$ and moreover, in view of the expected experimental sensitivity on the trilinear, not relevant. The diagram in the right of Fig.~\ref{tril} gives a 
momentum dependent correction of the same size $\sim 1/N$ as that induced by the wave function correction. This is also below
the quoted future sensitivities, modulo the unlikely possibility that
the momentum dependence of the correction could be used to boost up the signal.

\subsection{Dark matter constraints and simple fixes}
\label{sec:DM}

Within our framework $S$ is exactly stable. Unfortunately it cannot be identified with the dark matter because such a possibility appears to be in conflict with current direct and indirect dark matter searches. 

Let us see why this is the case. For our mechanism to work  $S$ must have been in thermal equilibrium 
at temperatures above the weak scale. This happens for 
 $|\lambda_{HS}|\gtrsim10^{-8}$, or equivalently $N\lesssim10^8(|\lambda_{HS}|N)\sim10^9$. Its present-day energy density $\rho_S$ is thus determined by standard freeze-out dynamics. The main number-changing processes setting $\rho_S$ are annihilations into SM particles controlled by the interaction $\lambda_{HS}H^\dagger HS^2$. In the regime of interest, $m_s> m_h/2$, the  Higgs propagator is always off resonance, and the largest annihilation rate is into Higgs  pairs (for $m_s>m_h$) and vector boson pairs (for $m_s>m_W$). The thermally averaged cross section is $\langle\sigma(S_iS_j\to{\rm SM})v\rangle\equiv\delta_{ij}\langle\sigma_{\rm ann}v\rangle$ with $\sigma_{\rm ann}v$ maximized above threshold by $\langle\sigma_{\rm ann, max}v\rangle\sim\lambda_{HS}^2/(8\pi m_s^2)$. The density $n_i$ of each of the $i=1,\cdots,N$ scalar components is determined solving the Boltzmann equation $dn_i/dt+3Hn_i\approx-\sum_j\langle\sigma(S_iS_j\to{\rm SM})v\rangle(n_in_j-n_i^{\rm eq}n_j^{\rm eq})$ with initial condition $n_i\sim n_i^{\rm eq}\sim T^3/\pi^2$ at high $T>m_s$. Because by symmetry this equation is the same for each $n_i$ one finds that the total number density $\sum_in_i=n_{\rm tot}=Nn_i$ satisfies 
\ba\label{BE}
\frac{dn_{\rm tot}}{dt}+3H(T)n_{\rm tot}\approx-\frac{\langle\sigma_{\rm ann}v\rangle}{N}[(n_{\rm tot})^2-(n_{\rm tot}^{\rm eq})^2], 
\ea
which essentially coincides with the evolution of the density for a single scalar dark matter candidate with an effective annihilation cross section $\langle\sigma_{\rm ann}v\rangle/N$. The solution of \eqref{BE} is a function $n_{\rm tot}(T)$ that decreases in time until the freeze-out temperature $T_{\rm fo}/m_s\sim 1/30-1/20$ that occurs when $n_{\rm tot}(T_{\rm fo})\sim NH(T_{\rm fo})/\langle\sigma_{\rm ann}v\rangle$. From that time on $S$ was out of equilibrium and $n_{\rm tot}(T)$ continued to decrease mainly due to the expansion of the universe. For all $T<T_{\rm fo}$ the energy density per unit entropy remains approximately constant and given by $\rho_S/s=m_sn_{\rm tot}/s=\sqrt{180/\pi} N(m_s/T_{\rm fo})/[\langle\sigma_{\rm ann}v\rangle M_{\rm Pl}\sqrt{g_*(T_{\rm fo})}]$, where $H(T)\sim\sqrt{g_*}T^2/M_{\rm Pl}$ and $g_*\sim100$ (at decoupling the large number of $S$ degrees of freedom contribute negligibly to the entropy).
%~\footnote{The numerical factor $\sqrt{180/\pi}$ is obtained keeping all the factors of order unity.} 
We can now compare the latter with the observed dark matter density, which itself is of order five times larger than that of baryons, i.e. $\rho_{\rm DM}/s\sim5m_p\eta_b$ with $m_p\sim1$ GeV the proton mass and $\eta_b\sim10^{-10}$ the baryon density per unit entropy. Combining everything together we arrive at
\ba\label{rhoDM}
\frac{\rho_S}{\rho_{\rm DM}}&\sim&\frac{N}{\langle\sigma_{\rm ann}v\rangle M_{\rm Pl}}\frac{1}{m_p\eta_b}\\\no
&\gtrsim&\frac{N}{\langle\sigma_{\rm ann, max}v\rangle M_{\rm Pl}}\frac{1}{m_p\eta_b}\sim \left(\frac{N}{50}\right)^3\left(\frac{m_s}{100~{\rm GeV}}\right)^2\left(\frac{10}{\lambda_{HS}N}\right)^2.
\ea

In many respects $S$ behaves analogously to the singlet scalar dark matter ($N=1$) studied extensively in the literature (see e.g.~\cite{Cline:2013gha,Beniwal:2015sdl,Athron:2017kgt} and references therein). For example, similarly to what we saw for the thermal abundance, the annihilation rates relevant to indirect dark matter searches are the same as those of a single scalar up to a suppression $1/N$. As a consequence, once the cross section for annihilation into SM particles at freeze-out is fixed to the value required to explain the dark matter of the universe, indirect signatures are effectively the same as for the standard $N=1$ case. The absence of observational evidence of such signatures already represents an important constraint on the model~\cite{Cline:2013gha,Beniwal:2015sdl,Athron:2017kgt}. 

However, when considering direct detection signals the situation is much worse compared to the well-studied singlet scalar dark matter model because in our scenario the number of expected events is increased by a factor $N\gg1$. 

Direct detection experiments set limits on the cross section for spin-independent scattering of $S$ off a nucleon $n$, which we write as $\sigma_{\rm DD}(S_in\to S_in)$ per each component $i$, and the signal is proportional to the total present-day number density: $Nn_i\sigma_{\rm DD}=n_{\rm tot}\sigma_{\rm DD}=(\rho_S/m_s)\sigma_{\rm DD}$. An explicit calculation gives~\cite{Cline:2013gha,Beniwal:2015sdl,Athron:2017kgt}
\ba
\sigma_{\rm DD}=\frac{\lambda_{HS}^2}{\pi m_s^2}\frac{\mu_{nS}^2m_n^2}{m_h^4}f_N^2, 
\ea
where $\mu_{nS}$ is the reduced mass of the $S$-nucleon system, $m_n$ the nucleon mass, and $f_N\sim0.3$ a nuclear form factor. Assuming $S$ constitutes the totality of the total dark matter, the current experimental bound for a singlet of mass not too far from $m_h$ roughly reads $\sigma_{\rm DD}<10^{-46}$ cm$^2$, see \cite{Aprile:2017iyp}. This translates into
\ba\label{DD1}
N\gtrsim2\times10^3,
\ea
which is much larger than required for $S$ to be the dark matter, see \eqref{rhoDM}. In our minimal scenario a large $N$ implies a small annihilation rate and consequently a larger abundance, and ultimately a signal stronger by a factor $\rho_S/\rho_{\rm DM}$ in direct detection experiments. More precisely, the direct detection cross section and the annihilation rate are related by $\sigma_{\rm DD}\sim({\mu_{nS}^2m_n^2}/{m_h^4})\langle\sigma_{\rm ann, max}v\rangle$. The actual bound in our model is $(\rho_S/\rho_{\rm DM})\sigma_{\rm DD}<10^{-46}$ cm$^2$ and reads:
\ba\label{DD}
10^{-46}~{\rm cm}^2>\frac{\rho_S}{\rho_{\rm DM}}\sigma_{\rm DD}\gtrsim N\frac{\mu_{nS}^2m_n^2}{m_h^4}\frac{1}{m_p\eta_bM_{\rm Pl}}\sim 
N\times10^{-44}~{\rm cm}^2,%~{\rm cm}^2<10^{-46}~{\rm cm}^2,
\ea
which is impossible to satisfy for any $N\geq1$ by at least a couple orders of magnitude.

We conclude that the minimal model (\ref{mod}) is not compatible with direct dark matter searches. In the reminder of this section we will propose three classes of extensions. We will entertain the possibility of adding new light particles that constitute the dark matter (Section~\ref{sec:ext1}), or breaking the stabilizing $O(N)$ symmetry, thereby making $S$ decay on cosmological scales and completely removing the dark matter from the model (Section~\ref{sec:ext2}), or finally introducing a new confining force (Section~\ref{sec:darkGlue}). The main lesson to be drawn from the following discussion is that it is possible to build simple extensions of the toy model (\ref{mod}) that are compatible with all observations. However, at this earlier stage we have no reason to prefer one of the following realizations over any other, so a detailed analysis of concrete scenarios is left for future work. It is nevertheless worth mentioning that the additional particles introduced in realistic models may affect the numerical analysis in Section \ref{sec:effV}, so that a proper model-dependent analysis of the finite-$T$ effective potential may be required. Furthermore, the richer structure  of the realistic models potentially makes the prospects of detection a bit more exciting compared to the original model of eq.~(\ref{mod}).

\subsubsection{Extension 1: adding a dark matter candidate}
\label{sec:ext1}

As our first extension we introduce new fermions that will play the role of the dark matter. For definiteness we promote $S$ to the traceless symmetric 2-index representation of a global $SO(n)$. With this slight modification the expression $S^2$ in the potential (\ref{mod}) should be interpreted as ${\rm tr}[S^2]$; in addition another quartic $\lambda_S'{\rm tr}[S^4]$ as well as a new soft trilinear $\mu_D{\rm tr}[S^3]$ are in principle allowed. With $\mu^2_Dn\ll 16\pi^2m_t^2$ (this condition may be enforced by an approximate $Z_2$ symmetry) and $\lambda_S'\ll\lambda_Sn$ the results of the previous sections, in particular the suppression of sphalerons at high $T$, are left unchanged provided we identify $N=n(n+1)/2-1$. 

Now, if no new ingredients are added, our estimate \eqref{rhoDM} shows that $\rho_S/\rho_{\rm DM}\gtrsim1$ for $N\gtrsim50$. Our plan here is to render $S$ unstable and let its decay products be the dark matter. This we achieve adding a pair of Weyl fermions $\chi$ with mass $m_\chi<m_s/2$ in the fundamental of $SO(n)$, and introducing the coupling $y\chi_i\chi_iS_{ij}$. Note that $\chi_i$ is exactly stable due to the $SO(n)$ symmetry and an accidental $Z_2$. In this new set up $S\to\chi\chi$ would quickly deplete the scalar population at a temperature $T_{\rm decay}\lesssim m_s$, provided at this scale the decay rate is comparable or larger than $H(T=T_{\rm decay})$. We will assume $T_{\rm decay}\gtrsim T_{\rm BBN}\sim1$ MeV to avoid affecting BBN. This corresponds to requiring $y^2\gtrsim4\pi\sqrt{g_*}{T^2_{\rm BBN}}/(m_s{M_{\rm Pl}})$.

With a sizable $y$, however, $\chi$ gets thermalized at $T\gtrsim m_s$ via reactions such as $\Gamma(\chi\chi\leftrightarrow SS), \Gamma(\chi\chi\leftrightarrow SSS)$, the latter being mediated by an off-shell $S$; having no lighter states to decay into, nor efficient annihilation channels, the present-day relic $\chi$ density would then be way too large to be identified with the dark matter. To ensure $\chi$ be a compelling dark matter candidate we impose the conservative conditions $\Gamma(SSS\to\chi\chi)\sim y^2\lambda_S^2n^3T/(4\pi)^5\lesssim H(m_s)$, $\Gamma(SS\to\chi\chi)\sim y^4n^2T/(4\pi)^3\lesssim H(m_s)$, which guarantee $\chi$ does not thermalize, and assume its primordial number density is negligible. In practice, for a coupling in the range
\ba\label{yPlanck}
4\pi\sqrt{g_*}\frac{T^2_{\rm BBN}}{m_sM_{\rm Pl}}\lesssim y^2\lesssim {\mathrm {min}}\left [4\pi\sqrt{g_*}\frac{m_s}{M_{\rm Pl}}\frac{\sqrt{N}}{\epsilon_S^2}, 4\pi\sqrt{\sqrt{g_*}\frac{m_s}{M_{\rm Pl}}\frac{4\pi}{{N}}}\right]
\ea  
$\chi$ was never thermalized at any $T<\Lambda$ and the present-day density is entirely controlled by $S\to\chi\chi$. The decay will generate a density of order $\rho_\chi/s\sim(\rho_S/s)(m_\chi/m_s)$, where the suppression of $m_\chi/m_s$ arises from the fact that the energy density of $\chi$ approximately scales as radiation from $T\sim m_s$ down to $T\sim m_\chi$, and only after that threshold $\chi$ started to behave as cold dark matter. Choosing the masses such that $(\rho_S/\rho_{\rm DM})(m_\chi/m_s)\sim1$ the fermion can thus be identified with the dark matter. The phase space bound $m_\chi\gtrsim$ keV~\cite{Tremaine:1979we} combined with \eqref{rhoDM} tell us that the initial $\rho_S$ cannot be arbitrarily large: the viable regime is in practice limited to $N\lesssim10^3-10^4$. The conflict with direct dark matter searches that the minimal model suffers from are here evaded: there is virtually no hope of detecting $\chi$ because all interactions between the dark matter and the SM are mediated by the Planck-suppressed coupling $y$, see \eqref{yPlanck}.

\subsubsection{Extension 2: breaking $O(N)$ softly}
\label{sec:ext2}

Our second class of extensions has an unstable $S$, there is no dark matter candidate and the direct/indirect detection constraints are trivially satisfied. 

Let us consider for instance the most general set of soft terms we can add to \eqref{mod}:
\ba\label{soft}
\delta V_{\rm soft}=a_i\frac{\mu^3}{g_S}S_i+b_{ij}\frac{\mu^2}{2}S_iS_j+c_{ijk}g_S\mu S_iS_jS_k+d_ig_S\mu S_i|H|^2,
\ea
where $\mu$ is a mass scale, $g_S$ a coupling and $a_i,b_{ij},c_{ijk},d_i$ arbitrary real matrices. We assume that $a_i,b_{ij},c_{ijk},d_i$ are of order unity and $\mu\ll |m_H|, m_S$, so that our analysis of the physics at finite temperature in the $O(N)$ symmetric model is not affected by the soft terms in \eqref{soft}. Our main task here is to identify what values of $\mu$ are large enough to trigger $S$ decay on cosmological scales and simultaneously small enough to be consistent with collider constraints. 

%It is fairly easy to meet these criteria because the new source of explicit $O(N)$ breaking can be naturally tiny. 

Note that $a_i,d_i$ are invariant under two (in general inequivalent) $O(N-1)\subset O(N)$, and when combined leave an $O(N-2)$ invariant. This means that in their presence only two components $S_i$ will be able to decay while additional soft terms are necessary to destabilize the remaining $N-2$ ones. Similarly, $b_{ij}$ leaves intact a $Z_2^N$ subgroup and is not enough by itself for our purpose either. Finally, $c_{ijk}$ breaks $O(N)$ completely. 

%When the group is completely broken, all the fields acquire VEVs of order $ d_i \mu g v^2/m^2_{S}$. 
The field fluctuations $\delta h$ and $\delta S_i=S_i-\langle S_i\rangle$ obtain a mass matrix that at leading order in $\mu/\Delta m$, where $\Delta m^2 \equiv m_h^2-m_s^2$, has the following structure
\ba
\label{massmat}
\sim\left(
\begin{matrix}
	m_h^2+\mathcal O (\mu^2) && d_i\mu g_S v\\
	d_i\mu g_S v && \delta_{ij}m_s^2+b_{ij}'\mu^2
\end{matrix}
\right),
\ea
where $b'_{ij}\mu^2=b_{ij}\mu^2 + 6 c_{ijk} g_S \mu \langle S_k\rangle$ with $\langle S_i\rangle\sim d_i \mu g v^2/m_s^2+{\cal O}(\mu^3)$. %After diagonalizing this matrix, the physical $\delta S_i$ combinations acquire a mixing with the Higgs. 
The diagonalization may be done in two steps: we first act with a rotation of angle $\sim d_i \mu g_S v / \Delta m^2$ to reduce \eqref{massmat} to a block diagonal form and then diagonalize the $N\times N$ block through a $O(N)$ orthogonal matrix $R$. Carrying out this procedure we find the mixing angles between $\delta S_i$ and the Higgs are given by
\ba
\sin\theta_i \sim R_{ij} d_j \frac{\mu g_S v}{\Delta m^2}.
\ea
Assuming that the order one coefficients $a_i$, $b_{ij}$, $c_{ijk}$ and $d_{i}$ are independent, we expect $b'_{ij}$ is also independent from $d_i$. Furthermore, we choose the coupling $g_S$ such that $(g_S v)^2\lesssim\Delta m^2$.~\footnote{If $(g_S v)^2 \gtrsim \Delta m^2$ the $b'$ contribution can become negligible compared to the off-diagonal terms in \eqref{massmat}. In this case the mass matrix would be nearly degenerate and the rotation $R$ would orient $d_i$ in the direction of a single component of the $\delta S_i$ fields, allowing only for its mixing with an angle $\sim\sqrt{N}{\mu g_S v}/({\Delta m^2})$, where we used the fact that a vector with random entries of order unity has $|d|\sim\sqrt{N}$. Since we want all the components of $S$ to decay, we choose to stay away from this region.} In this case {\emph{all}} physical components $\delta S_i$ have a non-vanishing mixing with the Higgs that allow for their decay into pairs of SM fermions or virtual gauge bosons. The decay width is of order $\Gamma_i\sim\sin^2\theta_i\Gamma_{\rm tot}$, where we used the fact that $m_{s_i}\sim m_h$ is a reasonable approximation and denoted by $\Gamma_{\rm tot}$ the Higgs width. In order not to spoil the successes of standard BBN cosmology we have to require all singlet components are removed from the plasma well before temperatures of order $T_{\rm BBN}\sim1\text{ MeV}$. Since $\Gamma_{\rm tot}\sim$ MeV this gives $\theta_i^2\gg\sqrt{g_*}{\rm MeV}/M_{\rm Pl}\sim10^{-21}$.

There is potentially another concern to take into account, however. If the scalars decay when their energy density dominate the universe, then a large entropy can be injected, in turn leading to a dilution of the baryon asymmetry. To avoid this we will conservatively require all singlet components decay before the temperature $T_*$ at which $\rho_S(T)=\rho_S(T_{\rm fo})(T/T_{\rm fo})^3$ starts to dominate over radiation. Recalling that in a standard cosmology matter-radiation equality takes place at around $T_{\rm eq}\sim1$ eV, using \eqref{rhoDM} and the benchmark values $\lambda_{HS}N=10$, $m_s=100$ GeV, we estimate
\ba\label{eq:Tstar}
T_*\sim T_{\rm eq}\left(\frac{N}{50}\right)^3\sim 1~{\rm MeV}\left(\frac{N}{5000}\right)^3.
\ea
For the values of $N$ considered in this paper $T_*<T_{\rm BBN}$, so this request is weaker than demanding BBN remains standard.

At the same time, bounds from collider experiments tell us that the mixing cannot be too large. In fact, too large mixing angles would imply the physical Higgs has couplings to the SM particles suppressed with respect to those predicted by the SM by a factor $\cos \theta$, with
\ba
\cos^2\theta \sim {1-\sum_i \sin^2 \theta_i}\sim 1 - {N}\left(\frac{\mu g_S v}{\Delta m^2}\right)^2.
\ea
To be consistent with experimental data, we require ${N}({\mu g_S v}/{\Delta m^2})^2\ll10\%$. 

The previous two conditions are satisfied in the region
\ba
10^{-21}\ll\left(\frac{\mu g_S v}{\Delta m^2}\right)^2\ll \frac{10^{-1}}{N}.
\ea
We see that with values $N\sim 100-1000$ there is a large portion of parameter space that allows us to remain compatible with both dark matter and collider bounds. 

\subsection{A more ambitious fix: confining $S$ and composite dark matter}
\label{sec:darkGlue}

In this section we will explore a scenario where $S$ is promoted to the anti-symmetric representation of a dark $SO(n)$ gauge group. Contrary to the previous model, the gauge symmetry here forbids a scalar trilinear and the renormalizable Lagrangian remains invariant under an accidental $Z_2$ symmetry:
\ba\label{Z2}
S\to -S.
\ea 
We will assume that the symmetry \eqref{Z2} is preserved by our UV completion at the scale $\sim\Lambda$. We do this in order to obtain a qualitatively different scenario compared to the one discussed in Section~\ref{sec:ext2}.~\footnote{In the absence of a $Z_2$ symmetry $S$ is expected to decay into exotic gauge bosons and SM particles. For example, taking $S$ in the 2-index symmetric of $SO(n)$ allows a trilinear coupling $S^3$, whose main effect would be to trigger a (1-loop) width into dark gluons at the renormalizable level. The $Z_2$-symmetric scenario discussed in this section reveals a richer phenomenology, which we find interesting to investigate.}

Our earlier discussion on symmetry non-restoration goes through qualitatively unchanged provided we identify $N=n(n-1)/2\sim n^2/2$. Yet, the calculation of the relic abundance of $S$ should take into account a new ingredient: the dark gauge bosons.

We estimate the population of $S$ at freeze-out observing that the dominant annihilation channel is into dark gauge bosons: 
\ba
\frac{\rho_S}{s}\sim\frac{8\pi m_s^2}{g_D^4(m_s)}\frac{m_s/T_{\rm fo}}{M_{\rm Pl}\sqrt{g_*(T_{\rm fo})}},
\ea
where the relevant number of relativistic degrees now should include the dark gluons as well, i.e. $g_*(T_{\rm fo})\sim n^2/2\sim N$. With $m_s\sim100$ GeV, a relic density comparable to the dark matter is obtained for $g_D^4(m_s)\sqrt{N}\sim10^{-3}-10^{-2}$. The present-day abundance of $S$ may be however significantly affected due to novel non-perturbative effects at $T\ll m_s$. %Before assessing whether $S$ can be a good dark matter candidate, however, we need to study what happens at late times. 
In fact, at temperatures of order the strong coupling scale, which we may estimate using the 1-loop beta function as $m_D\sim m_s\exp[-24\pi^2/(21g_D^2(m_s)(n-2))]\ll m_s$, the dark gluons and $S$ confine into $SO(n)$-singlet states with large self-couplings. It is therefore important to understand what implications this new phase can have on the relic density of the new particles. 

First of all, let us consider the bound states --- with masses $\sim m_D$ ---  that are formed by pure gluons, which we will refer to as dark glueballs. The CP-even dark glueballs are {{unstable}} and can decay into light SM particles via the effective interaction 
\ba\label{decayGlue}
\frac{g_D^2(n-2)}{96\pi^2}\frac{\lambda_{HS}}{m_s^2}H^\dagger HF_D^{\mu\nu}F_{D{\mu\nu}},
\ea
where $F^{\mu\nu}_D$ and $g_D$ are the field strength and coupling of the dark gauge group. Denoting by $\Phi$ the interpolating field for the CP-even dark glueballs, the decay may be estimated observing that at the confinement scale naive dimensional analysis --- combined with a large-$n$ counting --- suggests $g_D^2(m_D)n\sim16\pi^2$ and $F_D^{\mu\nu}F_{D{\mu\nu}}\sim m^3_D\Phi{n}/4\pi$. In practice the operator \eqref{decayGlue} interpolates a mixing of order $\theta\sim(\lambda_{HS}{\sqrt{N}}vm_D^3)/(4\pi m_s^2m_h^2)\ll1$ between the CP-even dark glueballs and the Higgs boson. The main decay channels the dark glueballs acquire are into SM fermions $f$, with a width of order $\Gamma({\rm dark\;glue}\to f\bar f)\sim\theta^2m_D(m_f/v)^2/(4\pi)$. 

We conservatively require the glueballs disappear as soon as they become non-relativistic. This request is motivated by the following consideration. At early times the universe expansion is controlled by dark radiation due to the large number $g_*\sim N$ of new degrees of freedom. By continuity we therefore expect the density of non-relativistic glueballs would dominate over SM radiation for a long time before they decay. If this was the case we would experience a significant dilution of $\eta_b$ due to either a large entropy production from their out-of-equilibrium decay or a prolonged period of dark glueball-dominated expansion. We evade these undesirable complications conservatively requiring the glueballs disappear as soon as they formed. This gives the order of magnitude bound $m_{D}\gtrsim (N/1000)^{1/5}(m_f/{\rm GeV})^{-2/5}$ GeV, which tells us that the dark dynamics should confine at a scale not too far from the GeV. The proximity between the required $m_{D}$ and the weak scale, or more precisely the $S$ mass, implies the new gauge symmetry should be rather strong already at $T\sim m_s$. Using our 1-loop estimate of $m_D$ this requirement approximately reads $g_D^2(m_s)n\sim3$.

The lightest CP-odd dark glueball is presumably heavier than the CP-even one, and can therefore annihilate into the CP-even states. The associated cross section times initial relative velocity (roughly of order $v_{\rm in}\sim\sqrt{T/m_D}\sim1$ at the temperatures relevant to this discussion) is $\sim\pi/[N^2m_D^2]$. The power of $N$ arises from the fact that trilinear (quartic) glueball couplings are of order $1/n$ ($1/n^2$), so the amplitude squared scales as $(1/n^2)^2\sim1/N^2$. For the typical parameters we are interested in, $m_D\sim1$ GeV and $N\sim10^3$ (see \eqref{DD1}), these processes should be efficient enough to deplete the CP-odd glueball abundance below the observed dark matter. The resulting relics would also eventually decay into SM particles via CP-violating loops involving the Higgs and SM fermions and gauge bosons. The time scale could be quite large, causing potential trouble at later stages of cosmological evolution (BBN, CMB, etc.). Without resorting to a detailed analysis, we notice however that all problems can be avoided by assuming the  topological vacuum angle $\bar\theta_{D}$ of the dark gauge group is non-vanishing. In that case the CP-odd glueballs  mix with angle $\bar\theta_{D}$ with the CP-even ones and decay on a time scale $1/[\bar\theta_{D}^2\Gamma({\rm dark\;glue}\to f\bar f)]\lesssim10^{10}[10^{-8}/\bar\theta_{D}]^2{\rm years}$.~\footnote{It is easy to convince oneself this additional source of CP violation does not affect baryogenesis: it becomes potentially relevant only after sphalerons have already shut off and no $B$-violating process is active.} 

%Despite whether hadronic $S$ states are or are not a dark matter candidate, there is one additional important constraint we have to take into account. 
The introduction of the dark gauge bosons opens a new decay channel for the Higgs; because we have seen that the dark glueballs typically have long lifetimes, this new channel corresponds to an invisible width for the Higgs and we have to make sure it is not too large. In the limit $m_s\gg m_h$, and neglecting radiative dark gluon corrections for simplicity, we obtain:
\ba\label{invglue}
\Gamma(h\to {\rm dark~gluons})=\left(\frac{g_D^2(n-2)}{16\pi^2}\right)^2\frac{m_h^3}{288\pi}\left(\frac{\lambda_{HS}v}{m_s^2}\right)^2n(n-1).
\ea
As an order of magnitude estimate, we will adopt the very same expression even in the realistic regime $m_s\sim m_h$. Taking as representative input values $m_s=100$ GeV, $|\lambda_{HS}|N=10$, and ${g_D^2(m_s)(n-2)}=3$ (as required to obtain a confinement scale around the GeV), we find that \eqref{invglue} is $<10\%$ of the total Higgs width as soon as $n\gtrsim20$, i.e. $N\gtrsim200$. %, which is interestingly close to the range of interest. 

\subsubsection{Reannihilation after confinement and Sglueball dark matter}
\label{rehannihilation}

Next, we want to understand what happens to the population of exotic scalars. $S$ confines into glueball-like states of mass $\sim m_s$ which belong to one of two classes of hadrons. %The present-day abundance of $S$ is then significantly reduced compared to its freeze-out value due to novel non-perturbative effects at $T\lesssim m_D$. We will not present a detailed analysis of this process, which requires dealing with complicated non-perturbative effects, but rather provide a qualitative picture of the physics involved. 
In the first class we find mesonic states, like ${\rm Tr}[SS]$, that carry no conserved $Z_2$ charge. These are Coulombian in nature, analogously to the quarkonium in QCD, have mass $2m_s-{\cal O}(\alpha_D^2m_D)$ and a characteristic size of order the Bohr radius $r_{SS}\sim1/[\alpha_Dm_s]$. Here $\alpha_D=g_D^2n/(4\pi)\sim0.2-0.3$ is the 't Hooft coupling renormalized at a scale of order $r_{SS}$ --- somewhere in between $1/m_D$ and $1/m_s$. The absence of any conserved charge implies they quickly decay, with rates $\sim\alpha_D^5m_s$ similarly to charmonium, essentially removing part of the original $S$'s from the plasma. The other class of $S$-hadrons is built out of an odd number of $S$'s, and carry a non-trivial charge~\eqref{Z2}. The stable configuration, which we will call {\emph{$S$glueball}}, is composed of one $S$ and gluons. It has mass $\sim m_s$ and binding energy $\sim m_D$. Its stability, however, does not immediately guarantee that the {\emph{$S$glueball}}s represent a significant fraction of the dark matter, as they can efficiently re-annihilate
at late times through the strong interaction process
 \ba\label{sglueRate}
{\rm{\emph{$S$glueball}} + {\emph{$S$glueball}}}\to~{ [SS]}+ {\rm glueballs}.
\ea
followed by the cosmologically fast decay of the mesons into glueballs.

To establish how many {\emph{$S$glueball}}s are left today we need to quantify how efficient the latter process is. A crude estimate can be obtained by a simple dimensional argument. At small $n$ the only available parameter is $m_D$, which controls
both the size of the   {\emph{$S$glueballs}} and the mass splittings. In this limit the cross section is thus expected to be geometrical $\sim \pi/m_D^2$. At large $n$ the trilinear coupling between two {\emph{$S$glueballs}} and unstable mesons $[SS]^*$ scales like $1/n$, while the spectrum is basically fixed. We thus expect that, as long as the resonant processes {\emph{$S$glueballs}}+{\emph{$S$glueballs}}$\to[SS]^*$ are kinematically accessible, the cross section simply features a reduction by $1/n^2$: $\sigma_{Sg}\sim \pi/(m_D^2 n^2)$. If the resonant reactions are not allowed, on the other hand, the generic large-$n$ expectation is that $\sigma_{Sg}\sim \pi/(m_D^2 n^4)$, corresponding to a two body ($[SS]$ + glueball) final state.

We can understand these claims  more quantitatively by considering the partial wave decomposition of the annihilation cross section
\beq\label{Landau}
\sigma_{fi}=\frac{\pi}{k_i^2}\sum _{\ell =0}^\infty (2\ell+1) |S_{fi}^{(\ell)}|^2
\eeq
where by $i$ we indicate the initial two {\emph{$S$glueball}} state, by $k_i$ their center of mass momentum and by $f$ any state they could annihilate into. At large $n$ we expect $S_{fi}^{(\ell)}$ to be well described by Breit-Wigner amplitudes mediated by narrow $[SS]^*$ meson states
\beq
|S_{fi}^{(\ell)}|^2\sim \sum_{r_\ell}\frac{\Gamma_i^{(\ell, r_\ell)}\Gamma_f^{(\ell, r_\ell)}}{(E_i - M^{(\ell, r_\ell)})^2+(\Gamma_{\rm tot}^{(\ell, r_\ell)}/2)^2},
\eeq
with $r_\ell$ any additional label identifying the intermediate states. The spectrum and the widths are controlled by $m_D$ and by $n$. In particular, the typical expectation  for the {\emph{$S$glueball}} and $[SS]^*$ masses is respectively 
 $M_{Sg}=m_s+c_{Sg}m_D$ 
and  $M^{(\ell, r_\ell)}=2m_s+c^{(\ell, r_\ell)} m_D$, with the $c$'s $O(1)$  coefficients. 
Given the initial energy $E_i=2\sqrt{M_{Sg}^2+k_i^2}$  the Breit-Wigner is thus controlled by
\beq
E_i-M^{(\ell, r_\ell)}\simeq(2c_{Sg}-c^{(\ell, r_\ell)}) m_D+ \frac{k_i^2}{M_{Sg}}\equiv \bar c^{(\ell, r_\ell)} m_D+ E_k
\label{EmenoM}
\eeq
with $E_k\sim T$ the kinetic energy of the initial {\emph{$S$glueball}} pair. On the other hand,
 $\Gamma_i^{(\ell, r_\ell)}\equiv\Gamma([SS]^*\to$ {\emph{$S$glueball}}+{\emph{$S$glueball}}$)$ and $\Gamma_f^{(\ell, r_\ell)}\equiv\Gamma([SS]^*\to[SS]$+{\emph{glueballs}}$)$, as well as the total width, are all of order $\Gamma\sim m_D/n^2$. 
Now, as explained below, reannihilation is dominantly taking place around $T\sim m_D/(20\div 30)$ so that the thermal breadth
of the energy denominator \eqref{EmenoM} is larger than the resonance width as soon as $n^2\gtrsim 30$. Thus, in our case, resonant annihilation happens for resonances that are within the thermal breadth of the initial state energy. Assuming resonances are spaced by $O(m_D)$  the {\it probability} for that to happen is  $p_r\sim T/m_D$. Such resonant partial waves give a contribution to the cross section 
\beq
\sim \frac{\pi}{k_i^2} (2\ell +1) \frac{\Gamma}{T}\sim  \frac{\pi}{k_i^2} (2\ell +1) \frac{m_D}{T}\frac{1}{n^2}
\label{res}
\eeq
while non-resonant partial waves give 
\beq
\sim  \frac{\pi}{k_i^2} (2\ell +1) \frac{\Gamma^2}{m_D^2}\sim  \frac{\pi}{k_i^2} (2\ell +1)\frac{1}{n^4}.
\label{nonres}
\eeq
To complete our estimate we should finally take into account that, given the initial momentum $k_i$ and given the typical
range $1/m_D$ of the interaction, we expect only partial waves up to $\ell_{max}\sim k_i \times 1/m_D$ to significantly  contribute.  A fraction $p_r\sim T/m_D$ of such waves will be resonant, so that, as long as $p_r\ell_{max}\gg 1$, we statistically expect a number $\sim \ell_{max} T/m_D$ of resonant channels. Combined with eq.\eqref{res}, this leads to an estimate
\beq\label{res1}
\sigma_{fi} \Big\vert_{p_r\ell_{max}\gg 1}\sim \frac{\pi}{k_i^2}\ell_{max}^2 \frac{1}{n^2}\sim \frac{\pi}{m_D^2}\frac{1}{n^2}\,.
\eeq
 On the other hand, for $p_r\ell_{max}\lesssim 1$ there is a good chance  no channel is resonant, so that, according to eq.\eqref{nonres} a more likely estimate
for  this case is
 \beq\label{nonres1}
\sigma_{fi} \Big\vert_{p_r\ell_{max}\lesssim 1}\sim \frac{\pi}{k_i^2}\ell_{max}^2 \frac{1}{n^4}\sim \frac{\pi}{m_D^2}\frac{1}{n^4},
\eeq
where the two powers of angular momentum arise from summing over all partial waves as indicated in \eqref{Landau}. The parameters of our scenario make  non-resonant annihilation much more plausible. Indeed $\ell_{max}\sim \sqrt{Tm_s}/m_D$ so that 
\beq
p_r\ell_{max}\sim \left(\frac{T}{m_D}\right)^{3/2}\left(\frac{m_s}{m_D}\right)^{1/2}\sim 0.1
\eeq
Nonetheless, without a complete non perturbative control of our model, it seems impossible to reach a definite conclusion.
We shall thus discuss both possibilities.

Having obtained a rough estimate of the cross section for \eqref{sglueRate} (see \eqref{res} and \eqref{nonres}) we can proceed to calculate the present-day abundance of {\emph{$S$glueball}}s. The latter is the result of a complicated system of coupled Boltzmann equations involving the densities of {\emph{$S$glueball}}s, mesons, and glueballs. Yet, we expect the dominant phase of $S$ depletion takes place at temperatures when the inverse reactions in \eqref{sglueRate} are negligible. A number of considerations suggest that the critical temperature $T_D$ at which this condition starts to hold lies in the range~\footnote{The upper bound may be estimated as follows. Because the $[SS]$ mesons decay instantaneously (in Hubble units) as soon as $T<m_D$, one may naively expect that the inverse reactions in \eqref{sglueRate} would be very unlikely already at $T<m_D$. %: the excited ones decay with rates $\sim m_D/N$ into the ground state that then quickly disappears. 
However, during meson de-excitation and decay many glueballs are injected in the bath, and these latter may occasionally find an unstable meson and dissociate it back into {\emph{$S$glueball}}s. This tells us that $T_D$ must be safely below the glueballs mass scale. A lower bound on $T_D$ may instead be identified observing that if we wait a bit longer, at $T<m_D/(20\div30)$, the glueballs have frozen out and the rate for the inverse \eqref{sglueRate} reaction is smaller than the expansion rate.} $m_D/(20\div 30)<T_D<m_D$. In the following we will therefore adopt the intermediate value $T_D=m_D/10$ as a reference. For $T<T_D$ the {\emph{$S$glueball}}s yield $Y$ approximately follows the equation $dY/dt=-s\langle\sigma_{Sg}  v_{\rm in}\rangle Y^2$. We found that $\langle\sigma_{Sg}  v_{\rm in}\rangle\sim\sqrt{T/m_s}\pi/(N^am_D^2)$, where $a=2$ generically and $a=1$ for the special case of resonant annihilation. The current energy density encapsulated in the {\emph{$S$glueball}}s is finally estimated as
\ba\label{depletion}\no
\frac{\rho}{s}\sim m_sY_0&=&\frac{m_sY_D}{1+Y_D\int^{t_0}_{t_D}{dt}~{s\langle\sigma_{Sg} v_{\rm in}\rangle}}\sim\frac{m_s}{\int^{t_0}_{t_D}{dt}~{s\langle\sigma_{Sg} v_{\rm in}\rangle}}\sim\frac{N^am_Dm_s}{{\pi}M_{\rm Pl}\sqrt{g_{*}(T_D)}}\frac{m_D}{T_D}\sqrt{\frac{m_s}{T_D}}\\
&\sim&10^{3(a-2)}\;\frac{\rho_{\rm DM}}{s}\left(\frac{N}{10^3}\right)^a\left(\frac{m_D}{1~\rm GeV}\right)^2\left(\frac{m_s}{100~\rm GeV}\right)\left(\frac{0.1~{\rm GeV}}{T_D}\right)^{3/2},
\ea
where we have taken the {\emph{$S$glueball}} mass to be $m_s$ and neglected their small binding energy ${\cal O}(m_D)\ll m_s$. The primordial {\emph{$S$glueball}}s yield at $T_D$, denoted by $Y_D$, %estimated to be $Y_D\sim8\pi m_s/[M_{\rm Pl}g_D^4(m_s)\sqrt{g_*(T_{\rm fo})}]$ with $g_D^2(m_s)n\gtrsim3$, is 
is typically a fraction of order unity of the freeze-out $S$ population. In the second line of \eqref{depletion} we used the fact that $\int^{t_0}_{t_D}{dt}~{s\langle\sigma_{Sg} v_{\rm in}\rangle}>1/Y_D$ for values of $N$ satisfying $N\ll10^8$. The main message to be qualitatively inferred from \eqref{depletion} is that the {\emph{$S$glueballs}} may be the dark matter. This conclusion may however be invalidated in the (unlikely) presence of a resonance (i.e. if $a=1$).

\subsubsection{An extension with mesonic dark matter}
\label{ssprime}

Here we show that there exists a minimal variation of the present model in which the issue of resonant vs non-resonant annihilation does not arise and the new scalars can be robustly argued to be the dark matter. 

We introduce another scalar $S'$ of mass $m_{s'}\sim m_s$ in the adjoint of $SO(n)$ preserving a $Z_2\times Z'_2$ symmetry, and argue that the $[SS']$ meson becomes a potential dark matter candidate. These are quarkonium-like states, with binding energy $\sim\alpha_D^2\mu_{SS'}$ and size $r_{SS'}\sim1/[\alpha_D\mu_{SS'}]$, where $\mu_{SS'}$ is the reduced mass of the $SS'$ system. Because of the sizable binding energy, $\alpha_D^2\mu_{SS'}> m_D$, the decay into a ``{\emph{$S$glueball}}+{\emph{$S'$glueball}}" pair is expected to be kinematically forbidden. As a consequence, the $[SS']$-meson is exactly stable. The population of our dark matter candidate is mainly governed by annihilation into pairs of unstable $SS, S'S'$ mesons (plus possibly dark glueballs):
\ba\label{mesons}
[SS']+[SS']\to [SS]+[S'S']
\ea
For $m_{s'}\neq m_s$ this channel is always energetically favored because the masses $M_{SS,S'S',SS'}$ of the mesons satisfy $\Delta M\equiv(M_{SS'}+M_{SS'})-(M_{SS}+M_{S'S'})\sim\frac{\alpha_D^2}{2}\frac{(m_{s}-m_{s'})^2}{m_{s}+m_{s'}}>0$. 
The reaction rate importantly depends on the final  phase space, which is simply determined by the 
 relative velocity of the mesons in the center of mass frame. Indicating by $\mu_{SS'}$ the reduced mass of the final state, in the non-relativistic limit we have
\ba\label{53}
v_{\rm fin}^2= v_{\rm in}^2\frac{M_{SS'}}{2\mu_{SS'}}+ 2\frac{\Delta M}{\mu_{SS'}}\simeq v_{\rm in}^2\frac{(m_s+m_{s'})^2}{4m_s m_{s'}}+
\alpha_D^2\frac{(m_{s}-m_{s'})^2}{m_{s}m_{s'}}[1+{\cal O}(\alpha_D^2)].
\ea
Of course we have $v_{\rm in}=v_{\rm fin}$ when $m_{s'}=m_s$, otherwise $v_{\rm fin}\sim \alpha_D$. The rate for \eqref{mesons} is dominated by the s-wave channel (the typical angular momentum is indeed $\ell\sim \mu_{SS'}v_{\rm in}r_{SS'}\sim v_{\rm in}/\alpha_D\ll1$). The cross section is therefore set by the Bohr radius of the initial states, $\sigma_D\propto (\pi r_{SS'}^2)(v_{\rm fin}/v_{\rm in})$, up to a ratio of final and initial relative velocities arising from the incoming flux ($\propto v_{\rm in}$) and the phase space ($\propto v_{\rm fin}$). In addition, since these are genuinely $2\to{2}$ reactions there is a $(1/n^2)^2$ factor from large-$n$ counting. Overall we obtain that the quantity relevant to our calculation, i.e. an average of the cross section times relative velocity of the initial states, is given by 
\ba
\langle\sigma_D  v_{\rm in}\rangle\sim\pi r_{SS'}^2\times \langle v_{\rm fin}\rangle\times\frac{1}{N^2}\sim\frac{\pi}{N^2\alpha_D\mu_{SS'}^2}\times
\begin{cases}
1 & m_{s'}\neq m_s \\
\frac{\langle v_{\rm in}\rangle}{\alpha_D} & m_{s'}= m_s. 
\end{cases}
\ea
This is a factor $m_D^2/(\alpha_D\mu_{SS'}^2)<1$ ($m_{s'}\neq m_s$) or $\langle v_{\rm in}\rangle m_D^2/(\alpha_D^2\mu_{SS'}^2)<1$ ($m_{s'}= m_s$) times the result \eqref{nonres} obtained above for the nonresonant {\emph{$S$glueball}} annihilation.  A calculation similar to \eqref{depletion}, which may be repeated here with $T_D\sim m_D$, then suggests that the stable $[SS']$ meson may be the totality of the dark matter for the typical choice $\mu_{SS'}\sim50$ GeV, $m_D\sim1$ GeV, $N\sim10^3$ (consistent with \eqref{DD1}), and finally $\alpha_D\sim0.2-0.3$. (Note that the term $Y_D$ in the denominator of \eqref{depletion} is not negligible if the abundance at freeze-out was already comparable to that of the present-day dark matter.) The above estimates are only qualitative, however. We plan to come back to this interesting topic in the future.

\vskip 1.0truecm
We have thus shown that the introduction of a dark gauge symmetry allows us to evade the dark matter constraints consistently with all current bounds. We should however warn the reader that, as opposed to the extensions discussed in Section \ref{sec:ext1} and \ref{sec:ext2}, these models introduce novel elements that can affect our analysis of the thermal effective potential. In particular, the sizable $g^2_D(m_s)n\sim3$ required to ensure a fast enough decay rate for the dark glueballs also generates a non-negligible positive contribution to the thermal $S$ mass and thus tends to suppress the negative corrections to $m_{h,0}^2(T)$. Simultaneously, a non-negligible gauge coupling helps pushing the Landau pole of $\lambda_S$ to higher scales, facilitating an extension of our models beyond $\Lambda>100$ TeV. While our earlier results on the effective Higgs potential at finite $T$ will remain qualitatively correct, a careful assessment of $N_{\rm min}$ should take these novel elements into account.

\section{Comments on alternative models}
\label{sec:alter}

We would like to argue now that the same qualitative conclusions we found for the $O(N)$ model extend to any large-$N$ realization of the low energy sector. Furthermore, it is possible to build small $N$ sectors by introducing new EW-charged scalars.

\subsection{Large-$N$ models} 

We have found that a large-$N$ dynamics offers a tractable framework in which a parametrically large Higgs VEV $h(T)/T$ can be obtained while maintaining perturbativity. It is a simple exercise to show that essentially all we have seen for the specific $O(N)$ model of \eqref{mod} generalizes to a wider class of nearly-conformal large-$N$ dynamics coupled to $H^\dagger H$. The interaction does not necessarily involve fundamental scalars, and could in principle be built out of a fermionic pair or a gauge field strength. In these latter cases it may thus be possible to relax the stability constraint, that has played a key role in our analysis, and even avoid the introduction of a new hierarchy problem if the new coupling to the Higgs is non-renormalizable, i.e. it has dimension $d\gtrsim4$. Unfortunately, in order to overcome the resulting suppression $(T/\Lambda)^{d-4}$ in the thermal Higgs mass --- and thus maintain the $B+L$ asymmetry down to low $T$ --- we would be forced to take extremely large values of $N$, which is in tension with the considerations made below \eqref{etab}. Overall, the option of a non-renormalizable coupling to $H^\dagger H$ seemed less convincing to us.

Even sticking to scenarios in which a renormalizable coupling to $H^\dagger H$ is built out of fundamental scalars, strictly speaking there is no necessity of having an approximate $O(N)$ symmetry: all is needed for our program to be realized is many degrees of freedom coupled to the Higgs mass operator. Consider for instance a model with $N$ scalars $S_i$  and  potential 
\ba\label{modifiedmod}
V_4=\lambda'_{HS}H^\dagger H\sum_iS_i^2+\frac{\lambda'_S}{4}\sum_iS_i^4+\lambda' \sum_{ij} S_i^2S_j^2,
\ea
where we may assume $\lambda'\sim{\lambda'}_{HS}^2/16\pi^2\ll\lambda'_S$ for simplicity. In this alternative model, while we still have $\delta m_H^2\sim\lambda'_{HS}NT^2/12$ at high $T$, the $N$ dependence of the constraints from stability and perturbativity are modified. Yet the analogue of eq.~\eqref{Nrough} giving the lower bound on $N$ is remarkably unaffected  with respect to the original model of eq.~\eqref{mod}.
Indeed the loop expansion parameter for the $S$ quartic is now given by $\epsilon_S'\equiv \lambda_S'/16\pi^2$ which does not carry any $N$ dependence. On the other hand considering the potential along the direction  $S_i^2\sim S_j^2$ one obtains the stability constraint $(\lambda'_{HS}N)^2<\lambda_H(\lambda'_SN)$. This second result can be conveniently written as
\beq
N\gtrsim\frac{(\lambda'_{HS}N)^2}{\lambda_H \lambda_S'}
\eeq
which leads to the same constraint of eq.~\eqref{Nrough}, once we notice that the loop expansion parameter is now $\epsilon_S'=\lambda_S'/16\pi^2$ rather than $\epsilon_S=\lambda_SN/16\pi^2$. The parameter  $\epsilon_S'$ also practically replaces
$\epsilon_S$  when computing the effective potential, 
the $S_i$  thermal masses in particular. 
So we again find no qualitative new feature arising from \eqref{modifiedmod}. This simple alternative shows that, in scenarios
where our mechanism is realized through couplings to the Higgs bilinear $H^\dagger H$, the lower bound on $N$ is structurally robust.

It would be interesting to investigate other applications for the large number of degrees of freedom coupled to $H^\dagger H$ that these models feature. A possible direction to explore is a potential connection with the scenario of~\cite{ArkaniHamed:2005yv}\cite{Arkani-Hamed:2016rle}, where $N\gg1$ scalars are introduced to address the cosmological constant problem and the hierarchy problem respectively. See also \cite{Cohen:2018cnq} for a connection to the dark matter problem.

\subsection{Models with smaller $N$} 
\label{sub:smallN}

It is possible to realize simple perturbative scenarios that support significantly smaller values of $N$. The key point is that we do not necessarily need to couple the new light degrees of freedom, say $S$, to the SM Higgs at all. Our mechanism would work equally well if we introduced a new EW-charged scalar that interacts with $S$ via a negative quartic, but with the SM mainly via the EW force. As a concrete example we may introduce an inert $H'$ at the weak scale, having the same SM charges as $H$ but carrying an unbroken $Z_2$ parity. It is the latter scalar that acquires a large VEV at finite $T$ and suppresses $B+L$ washout. Still, its zero-temperature mass squared may be chosen to be positive so that $H'$ does not contribute to EW symmetry breaking today, and all precision EW observables remain essentially unaffected. 

The discussion of Sections~\ref{firstapprox}, \ref{sec:T}, \ref{sec:effV} applies to this scenario as well provided we make a few important adjustments. First, the role of the Higgs boson is now played by $H'$, so its quartic $\lambda_{H'}$ is an unknown parameter. Second, the $Z_2$ symmetry forbids a tree-level interaction between $H'$ and the SM fermions, in particular the top quark. This eliminates the largest of the positive corrections in (\ref{honTrough}) and makes $A$ smaller. Furthermore, it implies the new quartic $\lambda_{H'}$ does not receive large renormalization effects and thus stays almost constant (or maybe increases a bit) up to the UV cutoff. The stability condition stays approximately unchanged through the RG scale, rather than becoming significantly stronger at high scales as it was in our model. Finally, the function $h'(T)/T$ now decreases with $T$, so the constraint $h'/T>1.2$ we derived using \eqref{gammab} becomes in fact conservative. All these effects tend to decrease the minimum $N$ required to avoid washout of the primordial asymmetry. For example, taking $\lambda_{H'}(m_t)=\lambda_H(m_t)/2$ and $\epsilon_S=0.01$ the rough estimate \eqref{Nrough} now gives $N\gtrsim60$. Repeating the numerical analysis of Section \ref{sec:effV} for $\lambda_{H'}(m_t)=\lambda_H(m_t)/2$, $\epsilon_S(m_t)=0.05$, and still imposing the conservative constraint $h'(T)/T>1.2$, we find that $N_{\rm min}\sim20$. This is a significant difference  compared to the $O(N)$ model considered in this paper. Less exotic scenarios may now be envisioned. For example, $N=24\gtrsim N_{\rm min}$ can be obtained by simply promoting $S$ to the adjoint representation of an $SU(5)$ gauge group, along the lines followed in Section \ref{sec:DM}.

The constraints discussed in Section \ref{sec:pheno} should be carefully re-assessed in models with additional EW-charged scalars. As concerns to our previous example, we note that the inert doublet $H'$ is typically a small fraction of the dark matter, and that current bounds from dark matter experiments are easily accommodated. Also, collider constraints are rather weak provided we choose the $H'$ couplings to the Higgs such that the mass splitting between its components is $\Delta\ll m_W$, but nevertheless sufficiently large to allow a fast decay of the charged scalar. This ensures the main signatures at colliders involve missing energy, that are poorly constrained (see e.g. \cite{Belyaev:2016lok}). Furthermore, with $\Delta\ll m_W$ the new contribution to the EW parameters is safely below current bounds.

\section{Electro-weak baryogenesis at $T_c\gg100$ GeV}
\label{sec:100TeV}

In this section we sketch two examples of physics for the EW phase transition occurring at the critical temperature $T_c\propto\Lambda$. Our main goal is to illustrate the feasibility of our program. In particular, focussing on our basic $O(N)$ model  we will show its compatibility  with a realistic scenario for baryogenesis at the scale $\Lambda$.

 \subsection{Weakly-coupled sectors}

We start by considering scenarios in which the dynamics at $\sim\Lambda$ is weakly-coupled. One simple option is to add a new scalar $\phi$ with potential couplings
\beq
\Delta V= \frac{m_\phi^2}{2} \phi^2+ \lambda_{H\phi}\phi^2 H^2 + \lambda_{\phi S} \phi^2 S^2 +\frac{\lambda_\phi}{4} \phi^4
\eeq
with 
 $m^2_\phi=\Lambda^2>0$,  $\lambda_{H\phi},\lambda_\phi>0$, $\lambda_{\phi S}<0$. As in the previous section, we can consistently assume $m_S^2(T)$ is always  positive, so that $\langle S\rangle =0$ at all temperatures. The main role of $S$ is to induce {\emph{negative}} thermal masses for the other two scalars: $m_{H,\phi}^2(T)\sim m_{H,\phi}^2-|\lambda_{HS,\phi S}|NT^2/12$.

\begin{figure}[t]
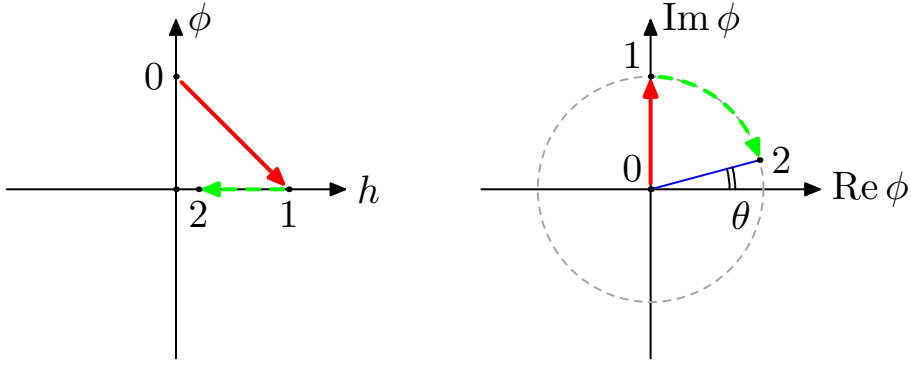

\begin{center}
\includegraphics[scale=1.5]{2StepPic}~~~~~~~~~~\includegraphics[scale=1.5]{NGBPic}
\caption{Schematic representation of the two kind of phase transitions studied in this section. The red arrow represents a first order phase transition at $T\sim T_c\propto\Lambda$, while the green dashed arrow shows a continuous change of the VEVs with the temperature $T<\Lambda$.
}\label{PhaseTransitionPics}
\end{center}
\end{figure}

To get the basic picture we can study the thermal vacuum dynamics in the same approximation used in Sec. \ref{firstapprox}, 
where the potential is determined by the quartic couplings  renormalized at $\mu  =T$ and by the  1-loop thermal masses. At high $T$, where $m_{H,\phi}^2(0)$ can be neglected,  and assuming $\lambda_{H\phi}>\sqrt{\lambda_H\lambda_\phi}$ one finds that there exist two distinct vacua separated by a barrier: ${\bf 0}\equiv(\phi\neq0,h=0)$ and  ${\bf 1}\equiv(\phi=0,h\neq0)$. We choose the parameters such that the deepest minimum at high $T$ is ${\bf 0}$, which simply requires
\ba\label{FirstOrder2}
\frac{m_\phi^4(T\gg T_c)}{\lambda_\phi}>\frac{m_H^4(T\gg T_c)}{\lambda_H}.
\ea
Defining $|\lambda_{HS}/\lambda_{\phi S}|\sqrt{\lambda_\phi/\lambda_H}=\zeta>0$ and assuming the $S$ loop tadpoles dominate the thermal masses, the above request is equivalent to $\zeta<1$. Because $m_\phi^2>0$ by assumption, as $T$ decreases $m_\phi^2(T)/T^2$ gets smaller in absolute value and eventually becomes positive when $T\leq T_\phi\sim12 m_\phi^2/(|\lambda_{\phi S}|N)$. At the critical temperature $T_c^2=T_\phi^2/(1-\zeta)[1+{\cal O}(m_H^2/m_\phi^2)]$ the solution ${\bf 1}$ starts to become deeper and a first order phase transition ${\bf 0}\to{\bf 1}$ can take place. For $T\ll T_c$ the physics is described by the model of Section \ref{sec:model}, and $h(T)$ continuously decreases with $T$ down to its zero-temperature value. A schematic picture of the phase diagram of this model is depicted in the left panel of Figure \ref{PhaseTransitionPics}.

%%%%%%%%%%%%%%%%%%
%%%%%%%%%%%%%%%%%%
%%%%%%%%%%%%%%%%%%
%%%%%%%%%%%%%%%%%% 

The transition ${\bf 0}\to{\bf 1}$ can be strongly first order, however, only if it is rather fast and occurs in the temperature range $T_\phi<T<T_c$. Indeed, after $T<T_\phi$ the fields can continuously roll down to $h\neq0$, thus avoiding the jump off the barrier. It is the strength of the couplings that controls the temperature $T_n$ at which bubbles of the true vacuum start to nucleate, and thus the onset of the phase transition, as well as its overall duration. While an accurate investigation is necessary to establish the region of parameter space in which EW baryogenesis actually occurs, we observe that one can in principle tune $1-\zeta\sim T_\phi^2/T_c^2\ll1$ in such a way that the barrier persists for a long time.~\footnote{Actually, with $1-\zeta\ll1$ the two vacua start almost degenerate at high $T$; hence a small drop in $T$ is enough for ${\bf 1}$ to dominate, i.e. $T_c\gg T_\phi\sim m_\phi$.} We are thus confident that a non-vanishing region of parameter space where $T_\phi< T_n<T_c$ can be engineered and our program realized.

 CP-violation at  the bubble walls, necessary to generate a baryon asymmetry during the transition, may be introduced in several ways. For  example one option is through vector-like fermions with SM quantum numbers,  whose mass matrix (with eigenvalues $\sim \Lambda$)  and Yukawa couplings to the Higgs jointly violate CP.
The key point is that these states are heavy enough to avoid constraints from the non-observation of rare flavor and CP violation at low energies.

\subsection{Strongly-coupled sectors}

It is interesting to investigate the alternative possibility that the heavy physics at $\Lambda$ be strongly-coupled. A seemingly natural assumption is then that both $H,S$ emerge as composite states. However, a more careful look reveals this might not be so simple to realize. Indeed, the required large  number of degrees of freedom in $S$ must be associated to a large number $N_{\rm fund}$ of fundamental constituents at the scale $\Lambda$. In such a framework a generic expectation is that all non-perturbative effects, including bubble nucleation, are exponentially suppressed by $e^{-{\cal O}(N_{\rm fund})}$. This would be a disaster for us, because it would mean that at $T = {\cal O}( \Lambda) $  bubble formation is too slow to drive a 1st order phase
transition. In this situation the phase transition would either never complete (see for instance \cite{Creminelli:2001th}), or complete at 
low temperatures where the CP violating sources are decoupled, or complete at the disappearance of the barrier as a smooth crossover. A more realistic possibility would be to have $H,S$  emerge as composites from two distinct dynamics: a ``small $N$" Higgs dynamics governing baryogenesis and a large-$N$ singlet dynamics designed to preserve the asymmetry.~\footnote{One possible way to couple the two sectors, and ultimately generate the desired coupling $\lambda_{HS}H^\dagger HS^2$ within a framework that does not suffer from the {\emph{big}} hierarchy problem, is to postulate the UV (small $N$) Higgs dynamics and the (large-$N$) $S$ dynamics possess scalar operators $O_{H,S}$ having scaling dimension close to two and being odd under some $Z_2$ symmetry. In that case the nearly marginal operator ${\cal L}_{\rm int}=\bar\lambda_{HS}O_HO_S$ would be allowed and, after confinement and symmetry braking at a scale $\Lambda$, would interpolate $\lambda_{HS}H^\dagger HS^2$.}

We here consider a simplified picture in which the Higgs is a composite Nambu-Goldstone boson of a new strong dynamics at a scale $\Lambda= g_H f$, whereas $S$ is taken to be elementary. We will argue that under reasonable assumptions, combining this picture with our scalar $S$ with a negative coupling to the Higgs results in a strongly first order EW phase transition. 

The picture is as follows. At $T>T_c$ the EW symmetry is unbroken. At $T=T_c$ the strong Higgs dynamics undergoes a first order phase transition associated to a symmetry breaking pattern ${\cal G}\to {\cal H}$. For definiteness we will have in mind the minimal $O(5)/O(4)$ scenario,~\cite{Agashe:2004rs} but our results straightforwardly generalize to more complicated cosets. The Higgs is a Nambu-Goldstone mode of the coset ${\cal G}/{\cal H}$, that we parametrize via $s_h\equiv\sin(h/f)$. For $T\leq T_c$ the composite Higgs is an exact flat direction of the strong dynamics. However, couplings to the SM and to $S$ break ${\cal G}$ explicitly, thus lifting such a degeneracy. Under the assumption that the largest $T$-dependent effects at temperatures $T\leq T_c$ are controlled by the ${\cal G}$-breaking coupling $\kappa$ between $S$ and the Higgs --- as in our toy model \eqref{mod} --- the full effective potential will typically acquire the form
\ba\label{CHT}
V(h,s=0)=\frac{m_h^2f^2}{8\xi}\left[-2\xi s_h^2+s_h^4+{\cal O}(s_h^6)\right]-\frac{\kappa}{4} NT^2f^2s_h^2\left[1+{\cal O}(1/N)\right],
\ea
where we used the fact that the couplings are engineered such that $\xi\equiv\langle s^2_h(T=0)\rangle\ll1$, as usual in composite Higgs models. (In \eqref{CHT} we assumed the coupling between the Nambu-Goldstone Higgs and $S$ is parametrized by $\sim\kappa s_h^2f^2S^2$, but one may consider more general options.)

The vacuum solution derived from \eqref{CHT} reads
\ba\label{CHsol}
\langle s^2_h(T)\rangle=
\begin{cases}
1 & T_*<T<T_c\\
\xi+(1-\xi)\frac{T^2}{T_*^2} & T<T_*,
\end{cases}
\ea
where $T_*^2=(1-\xi)m_h^2/(\kappa N\xi)$. We see that $S$ destabilizes the EW-symmetric vacuum $h^2=0$ (solution ${\bf 0}$) as soon as the phase transition occurs, so that the EW symmetry abruptly goes from unbroken to broken right at the critical temperature, where $h^2\sim f^2$ (solution ${\bf1}$), before approaching its zero-temperature value $h^2\ll f^2$ (solution ${\bf2}$). Specifically, the $W$-mass $m_W^2(T)=g^2f^2\langle s^2_h(T)\rangle/4$ {\emph{jumps}} from $m_W=0$ at $T>T_c$ to some non-zero $m_W(T)$ at all $T<T_c$. A schematic picture of the phase diagram of the present scenario, ${\bf 0}\to{\bf1}\to{\bf2}$, is shown in the right panel of Figure \ref{PhaseTransitionPics}.

The sphaleron shut-off condition \eqref{strongPT} in the present scenario becomes $f^2\langle s^2_h(T)\rangle/{T^2}\gtrsim1$. Using \eqref{CHsol} we find that this is satisfied for all $T_*<T<T_c$ provided the transition in the exotic dynamics is strongly first order, in the sense that $T_c\lesssim f$. To ensure $B+L$ preservation at $T<T_*$ we should instead require
\ba
\frac{f^2}{T^2}\langle s^2_h(T)\rangle=\frac{f^2}{T^2}\xi+\frac{\kappa N}{2\lambda_H}\gtrsim1.
\ea
Recalling that \eqref{CHT} was derived in the simplified limit $\kappa N\gg y_t^2$ in which the SM thermal loops are neglected, the latter condition is simply a re-writing of the request that \eqref{honTrough} be larger than unity.

To realize EW baryogenesis at $T_c\lesssim f$ we further postulate the existence of sizable CP-violating couplings for the Higgs at those temperatures. These necessarily arise from the new interactions that are introduced in order to reproduce the Yukawa couplings of the SM. For example, in modern incarnations of the Composite Higgs, these emerge via interactions like $yq{\cal O}_\Psi$, where $q$ is a SM fermion and ${\cal O}_\Psi$ a family of composite fermionic operators of the strong Higgs dynamics (see, e.g.~\cite{Contino:2003ve}\cite{Agashe:2004rs}). After symmetry breaking ${\cal O}_\Psi$ interpolate heavy vector-like resonances of mass $\Lambda=g_Hf$ and the coupling generates a $q/\Psi$ mixing, which ultimately results in the SM Yukawas. The phases in the coupling between $q,\Psi$ and the bubble wall, $\sim yf/\Lambda=y/g_H$, are unsuppressed at $T\sim T_c$, so EW baryogenesis can take place efficiently. On the other hand, all corrections to low energy rare processes are controlled by powers of $E/\Lambda\ll1$ (where $E$ is the characteristic energy) and can be within current bounds given the large $\Lambda$ considered here (see \cite{KerenZur:2012fr}\cite{Frigerio:2018uwx} for details).

\section{Conclusions}
\label{sec:discussion}

The absence of any indication of CP-violation beyond the SM puts significant pressure on standard realizations of EW baryogenesis. Here we have demonstrated that it is possible to build unconventional scenarios where all the relevant dynamics --- CP-violation and a strongly first order EW phase transition --- takes place at a new threshold $\Lambda\gg m_W$. The physics threshold $\Lambda$ may well be associated to the fundamental scale of flavor violation and safely lie in the range $\Lambda\sim100-1000$ TeV. The emerging picture suggests a connection between flavor and CP violation, the EW phase transition, and baryogenesis, and might feature interesting correlations in the corresponding indirect signatures (rare flavor- and CP-violating processes, primordial gravitational waves) all being characterized by $\Lambda\gg m_W$.

This new class of scenarios for EW baryogenesis rests on the existence of a new sector at the weak scale whose defining role is to prevent washout of the primordial $B+L$ asymmetry. Our setup essentially removes the CP-problem of the ordinary scenarios for EW baryogenesis, but retains their predictivity: there must exist new physics at the weak scale. The key novelty is that such new physics, i.e. our sector at the Fermi scale, can be SM-neutral and CP-conserving, and therefore more easily compatible with observations. All CP-odd phases beyond the SM are instead associated to physics at $\Lambda$ and decouple from low energy experiments, in agreement with observation. This setup paves the way to a multitude of novel realizations of EW baryogenesis. In Section \ref{sec:100TeV} we have sketched two concrete scenarios, in which the low-energy sector is combined with a complete picture at $\Lambda$ that includes CP-violation and a strongly first order EW phase transition, but many other realizations may be considered.

We have studied in detail a specific low-energy model consisting of a single scalar field in the fundamental representation of a new global $O(N)$, see (\ref{mod}), meant to represent an existence proof of our low-energy sector. The parameter $N$ is constrained by the requirement of perturbativity and stability of the EFT. Both constraints are significantly exacerbated by the largeness of the top quark Yukawa and the significant RG running of the Higgs quartic, and at the end of the day they force $N$ to be above $100$. We have seen, however, that such values may be dramatically reduced when non-minimal Higgs sectors at the weak scale are considered, see Section \ref{sec:alter}.

Our $O(N)$ model reveals an interesting $N\to\infty$ scaling in which qualitatively important effects are present at finite $T$ despite the fact that the zero-temperature dynamics is arbitrarily weakly-coupled. We studied in detail the effective potential at finite $T$ resulting from (\ref{mod}) and identified the regions of parameter space where our program is realized. Our large-$N$ expansion has been employed in order to obtain a reliable, and systematically improvable, approximation of $V_{\rm eff}$. 

The minimal model (\ref{mod})  also has an exactly stable dark matter candidate that is in conflict with current direct detection experiments. We have illustrated various   extensions where this bound is evaded without affecting our main conclusions. 
A particularly appealing option appears to arise when  gauging a subgroup of the global symmetry  of the minimal model. We have discussed a specific incarnation with gauged $SO(n)$ and with matter consisting of just a real scalar $S$ in the adjoint representation. We have identified a  range of parameters that is interesting for baryogenesis, dark matter, and late cosmology (BBN and CMB): $n$ is somewhat large $n\gtrsim 50$ and confinement happens at a scale of order $1$ GeV, parametrically well below the mass of $S$, $m_s\sim 100$ GeV, see Section \ref{sec:darkGlue}. Using generic properties of confining large-$N$ gauge theories we have argued that a bound state made of  one $S$ and gluons, the {\emph{$S$glueball}}, 
could be a plausible dark matter candidate. Besides the standard perturbative calculable  freeze-out  of the scalar $S$, which happens prior
to confinement, the large-$N$ suppression of rehannihilation after confinement  plays crucial a role. The resulting novel scenario for dark matter
is certainly worth a more detailed and careful study.

\section*{Acknowledgments}

We thank R. T. D'Agnolo, I. Baldes, D.E. Kaplan, S. Rajendran, M. Ramsey-Musolf, L. Senatore, for interesting discussions. A special acknowledgment goes to P. Meade for crucially reminding us of the possibility of EW symmetry non-restoration, G. Servant  for discussions that motivated us to develop this idea, M. Redi for important observations on the dark matter models of Section \ref{sec:darkGlue}, M. Shaposhnikov for explaining to us various key aspects related to EW baryogenesis, and A. Wulzer for comments on collider phenomenology. A preliminary version of the results was presented by LV at the MITP program ``Probing Baryogenesis via LHC and Gravitational Wave Signatures" in June 2018, and by AG at the 2018 Carg\`ese Summer School ``Mass: from the Higgs to Cosmology'' in July 2018. During the final stages of our work a few related papers appeared on the arXiv: refs.~\cite{Meade:2018saz,Baldes:2018nel}, which are based on  similar ideas, and \cite{Contino:2018crt}\cite{Gross:2018zha}, which have some overlap with our dark matter models of Section \ref{sec:darkGlue}. We thank R. Contino, A. Mitridate, A. Podo, M. Redi for discussions on their recent paper \cite{Contino:2018crt}. The Swiss National Science Foundation partially supported the work of A.G. and R.R. under contracts 200020-169696, and the work of L.V. under the Sinergia network CRSII2-16081.

\appendix

\section{Proof of $\langle S\rangle=0$}
\label{app:s=0}

In this appendix we demonstrate that the true vacuum of \eqref{mod}, subject to \eqref{muS}, satisfies $s^2=\langle S\rangle^2=0$. We will work at leading non-trivial order in $1/N$. Because $s$ is not a flat direction, subleading $1/N$ effects cannot impact the conclusion qualitatively. 

To prove our claim, let us first simplify our discussion by considering an ideal model without the Higgs, i.e. set $H=0$ in \eqref{aux1}. In this case the effective potential, that in general depends on the space-time independent value $s^2$, is simply obtained integrating out $\sigma$ from ${\cal L}_{\rm eff}=-\frac{1}{2}(m_S^2+\sigma)s^2+\frac{1}{4\lambda_S}\sigma^2+N\Gamma[m_S^2+\sigma]+{\cal O}(1/N)$. As already emphasized in the text, the large-$N$ approximation corresponds to an expansion in loops of the auxiliary field. The leading diagrams contributing to the effective potential for $s$ are therefore found solving the classical equation of motion of $\partial {\cal L}_{\rm eff}/\partial\sigma=0$. The effective potential for $s^2$ at leading order in $1/N$ is finally given by plugging the space-time independent classical solution $\sigma_c(s^2)$ back in $V_{\rm eff}(s)\equiv-{\cal L}_{\rm eff}(s,\sigma_c(s^2))$. By taking the total derivative of $V_{\rm eff}$ with respect to $s$ we see that
\ba\label{0V}
\frac{dV_{\rm eff}}{ds}=\left(\frac{\partial}{\partial s}+\frac{\partial\sigma_c}{\partial s}\frac{\partial}{\partial \sigma_c}\right)V_{\rm eff}=\frac{\partial}{\partial s}V_{\rm eff}=(m_S^2+\sigma_c)s.%\\\no&\equiv& m_{S, {\rm eff}}^2s,
\ea
Superficially, eq.~\eqref{0V} tells us that ${dV_{\rm eff}}/{ds}=0$ has two possible solutions: $s=0$ or $m_S^2+\sigma_c=0$. However, it turns out that $m_S^2+\sigma_c=0$ is unphysical. Indeed, in this latter case (and after having properly renormalized $m_S^2$) we have $-\partial {\cal L}_{\rm eff}/\partial\sigma_c=s^2/2+m_S^2/(2\lambda_S)-N\Gamma'[0]=s^2/2+m_S^2/(2\lambda_S)+NT^2/24=0$, which cannot be satisfied under the hypothesis \eqref{muS}. We thus conclude that $s=0$ is the only stable solution at this order. Also, differentiating twice the effective potential one finds that ${d^2V_{\rm eff}}/{ds^2}(s=0)\equiv m_S^2(T)=m_S^2+\sigma_c(s=0)\sim m_S^2+\lambda_SNT^2/12$ represents the full thermal mass. The direction $s$ is not flat and next to leading perturbative ${\cal O}(1/N)$ corrections will not spoil our conclusion.

An analogous result extends to the case with the Higgs field (and the full SM) included. The basic reason is that the SM loops are truly perturbative, and in particular not enhanced by powers of $N$, and can thus be neglected at leading order. As it was argued regarding the $1/N$ corrections, their inclusion cannot alter our arguments qualitatively. This observation allows us to simplify our analysis by considering the Lagrangian (see \eqref{aux1}):
\ba\label{treeApp}
{\cal L}_{\rm eff}&=&-\frac{1}{2}\left(m_H^2+\frac{\lambda_{HS}}{{\lambda_S}}\sigma\right)h^2-\frac{\lambda_H}{4}\left(1-\frac{\lambda_{HS}^2}{\lambda_H\lambda_S}\right)h^4\\\no
&-&\frac{1}{2}(m_S^2+\sigma)s^2+\frac{1}{4\lambda_S}\sigma^2+N\Gamma[m_S^2+\sigma]+{\cal O}(1/N, {\rm SM~loops}).
\ea
It then immediately follows that \eqref{0V} remains approximately correct. Furthermore, the equation of motion at $\sigma_c=-m_S^2$ is now given by $\lambda_Ss^2+m_S^2+\lambda_SNT^2/12+{\lambda_{HS}}h^2=0$: replacing the solution $h^2$, found differentiating \eqref{treeApp}, one can easily verify that this is incompatible with eq \eqref{muS} at any $T$. This again ensures that $s=0$ is the only consistent solution.

\section{Thermal potentials and beta functions}
\label{app:V}

Within a 1-loop approximation, the integration of a particle of spin $j=0,1/2,1$ and (field-dependent) mass-squared $M^2$ results in the following contribution to the finite-$T$ effective potential:
\ba\label{VT}\label{Vj}
V_{j}(M^2)&=&(-)^{2j}\frac{1}{64\pi^2}(M^2)^2[\ln(M^2/\mu^2)-c_j]\\\no
&+&(-)^{2j} T\int\frac{d\vec p}{(2\pi)^3}~\ln\left[1-(-)^{2j}\exp\left(-\frac{1}{T}\sqrt{{\vec p^2}+M^2}\right)\right].
%\\\no&=&T^4\left[-\frac{\pi^2}{90}+\frac{1}{24}y-\frac{1}{12\pi}y^{3/2}+{\cal O}(y^2)%-\frac{1}{64\pi^2}y^2\left(\ln\frac{y}{16\pi^2}-\frac{3}{2}+2\gamma_E\right)+\cdots\right].
\ea
We expressed \eqref{Vj} in the ${\overline{\rm MS}}$ scheme, with $\mu$ the renormalization point; $c_{0,1/2}=3/2$ for scalars and fermions whereas $c_1=5/6$ for vectors. 

An explicit 1-loop computation shows that (still in the ${\overline{\rm MS}}$ scheme):
\ba\label{betafunctions}
8\pi^2\mu\frac{d\lambda_H}{d\mu}&=&(8+N_H)\lambda_H^2+N\lambda_{HS}^2+{\cal O}(g^2\lambda_H,y_t^2\lambda_H,g^4,y_t^4)\\\no
8\pi^2\mu\frac{d\lambda_S}{d\mu}&=&(8+N)\lambda_S^2+N_H\lambda_{HS}^2\\\no
8\pi^2\mu\frac{d\lambda_{HS}}{d\mu}&=&\lambda_{HS}\left[(2+N_H)\lambda_H+(2+N)\lambda_S+4\lambda_{HS}+{\cal O}(g^2,y_t^2)\right],
\ea
where $N_H=4$. These equations were used in Section \ref{sec:effV} as a non-trivial check of the consistency of our effective potential \eqref{VeffT1}. In fact we verified that, generalizing $V_{\rm eff}$ to include $s\neq0$, \eqref{betafunctions} follow from the RG invariance of the effective potential.

\end{document}